\documentclass[12pt]{article}
\usepackage{amsfonts}
\usepackage[russian]{babel}
\title{
Гауссовские состояния для квантовых систем 
с линейными связями и квадратичными операторами Гамильтона
}
\author{ О.Ю.Шведов ,
\\
Россия, 119992, Москва, Воробьевы горы, \\
МГУ, физический факультет;
shvedov@phys.msu.ru
}
\date{}
\begin{document}

\maketitle
\footnotetext{Работа выполнена    
при   финансовой   поддержке   Российского   фонда
фундаментальных исследований, проекты 08-01-00601-а и 
05-01-02807-НЦНИЛ.}

\setcounter{page}{0}

\begin{flushright}
arXiv:0812.4629
\end{flushright}

\section*{Abstract}
В теоретической физике часто встречаются квантовые системы со связями:
все современные реалистичные модели квантовой теории поля относятся  к
этому типу.   Исследуются  общие  математические  свойства  простейших
квантовых систем  с  линейными  связями,  для   которых   гамильтониан
квадратичен по операторам умножения на координаты и дифференцирования.
Такие системы возникают при применении квазиклассического  приближения
к квантовым  системам  как  со  связями,  так  и без связей.  В работе
приводятся различные конструкции  гильбертова  пространства  состояний
системы со   связями,   исследуются  свойства  гауссовских  функций  и
комплексного ростка  Маслова;  приводится  обобщение  на  системы   со
связями теоремы   Маслова   о  взаимосвязи  устойчивости  классических
гамильтоновых систем и существовании гауссовской собственной функции у
квадратичного гамильтониана.
\newline
{\it Библиография:} 13 наименований.

{\it Ключевые слова:} комплексный росток Маслова, гамильтоновы системы
со связями, квантование.

\def\u#1#2{{\underset {#2} \to {#1} }}

\def\r#1{(#1)}
\def\beb#1\l#2\eeb{$$\begin{array}{c} #1 \end{array} \eqno({#2}) $$}
\def\bey#1\eey{$$\begin{array}{c} #1\end{array}$$}
\def\beq#1\l#2\eeq{$$ #1  \eqno({#2}) $$}
\def\bez#1\eez{$$  #1 $$}
\def\c#1{[#1]}
\def\i#1{#1.}
\def\Cal{\cal}
\def\Bbb{\mathbb}
\def\matrix #1 \endmatrix {\begin{array}{c} #1 \end{array} }
\def\matrixx #1 \endmatrixx {\begin{array}{cc} #1 \end{array} }

\newpage

\section{Введение}

Гамильтоновы системы  со  связями [1]  играют  важную  роль  в современной
теоретической физике.  В классической механике таких  систем  основным
объектом --- фазовым пространством [2] --- является не ${\Bbb R}^{2n} =
\{ (P,Q) | P,Q \in {\Bbb R}^n \}$,  а поверхность в  ${\Bbb  R}^{2n}$,
задаваемая $k$ уравнениями
\beq
\Lambda_a(P,Q) = 0, \qquad a = \overline{1,k}.
\l{1.1}
\eeq
Преобразование эволюции для системы со связями по форме не  отличается
от эволюции  механической системы без связей:  оно переводит начальное
условие для системы Гамильтона
\beq
\frac{dP_i(t)}{dt} =
- \frac{\partial H}{\partial Q_i}
(P(t),Q(t));
\qquad
\frac{dQ_i(t)}{dt} =
\frac{\partial H}{\partial P_i}
(P(t),Q(t)),
\l{1.2}
\eeq
лежащее на  поверхности  \r{1.1},  в  решение этой системы в момент $t$,
которое предполагается также лежащим на  этой  поверхности.  При  этом
$H(P,Q)$ называется функцией Гамильтона.

В случае, если скобки Пуассона
\bez
\{ A; B\} \equiv \sum_i
\left(
\frac{\partial A}{\partial Q_i}
\frac{\partial B}{\partial P_i}
-
\frac{\partial A}{\partial P_i}
\frac{\partial B}{\partial Q_i}
\right)
\eez
связей $\Lambda_a$  друг  с  другом  и  с  функцией   Гамильтона   $H$
обращается в нуль {\it на поверхности связи \r{1.1} }
\beq
\{ \Lambda_a; \Lambda_b \} = 0; \qquad
\{ \Lambda_a; H \} = 0,
\l{1.3}
\eeq
говорят о  системе со связями первого рода по Дираку.  Именно к такому
типу и относятся калибровочные теории [3,4] (в  частности  электродинамика).
Отметим, что  для системы со связями \r{1.1},  удовлетворяющими условиям
\r{1.3}, функции $\Lambda_a$ генерируют симплектические  преобразования,
задающие отношение    калибровочной    эквивалентности    на   фазовом
пространстве.

Рассмотрим теперь  квантовую  механику  систем  со  связями.  Основным
понятием квантовой  теории является гильбертово пространство состояний
$\Cal H$;  чистые состояния квантовой системы в каждый момент  времени
задаются элементами этого пространства [5,6]. Для системы с $n$ координатами
и $n$ импульсами в  качестве  $\Cal  H$  выбирают  $L^2({\Bbb  R}^n)$
волновых функций $\psi(\xi_1,...,\xi_n)$, зависящих от $n$ аргументов;
преобразование эволюции задается однопараметрической группой унитарных
операторов $e^{-i\hat{H}t}$,  которые  переводят начальное условие для
уравнения Шредингера
\beq
i \frac{d\psi(t)}{dt} = \hat{H} \psi(t), \qquad \psi(t) \in {\Cal H}
\l{1.4}
\eeq
в решение  этого  уравнения  в момент $t$.  Генератор группы $\hat{H}$
называется оператором   Гамильтона.   Формально   он   получается   из
классической функции   Гамильтона  заменой  классических  импульсов  и
координат $P_i,Q_i$ на квантовые импульсы и координаты --- операторы
\beq
\hat{p}_j = - i \frac{\partial}{\partial \xi_i};
\qquad
\hat{q}_j = \xi_j.
\l{1.5}
\eeq
Формально $\hat{H} = H(\hat{p},\hat{q})$.

Известны различные способы учета связей  в  квантовой  теории.  Первый
способ основан на модификации скалярного произведения [7]. Вместо обычного
скалярного произведения в пространстве функций
\bez
(\psi,\psi) = \int d\xi_1 ... d\xi_n
\psi^*(\xi_1,...,\xi_n) \psi(\xi_1,...,\xi_n)
\eez
выбирается другое:
\bez
((f,f)) = (f,\prod_a \delta(\hat{\Lambda}_a) f),
\eez
где $\hat{\Lambda}_a$   ---   операторы   связей,   равные   формально
$\hat{\Lambda}_a = \Lambda_a(\hat{p},\hat{q})$.  При этом говорят, что
волновая функция  $f(\xi_1,...,\xi_n)$ калибровочно эквивалентна нулю,
если $((f,f))=0$.  Пространство  функций  факторизуется   по   данному
отношению эквивалентности,   после   чего  рассматривается  пополнение
факторпространства. В   качестве   преобразования    эволюции    также
выбирается однопараметрическая    группа   $e^{-i\hat{H}t}$,   которая
сужается на факторпространство и расширяется на его пополнение.

Другой способ  учета  связей,  предложенный  Дираком [1],  заключается   в
наложении дополнительного условия
\beq
\hat{\Lambda}_a F = 0
\l{1.6}
\eeq
на функции    $F(\xi_1,...,\xi_n)$    из    пространства    состояний.
Преобразование эволюции также имеет  вид  $e^{-i\hat{H}t}$.  В  данном
способе квантования весьма нетривиально записать явную формулу для
скалярного произведения.

Приведенные способы учета связей рассматривались в простейшем абелевом
случае, когда
\beq
[\hat{\Lambda}_a; \hat{\Lambda}_b] = 0;
\qquad
[\hat{\Lambda}_a; \hat{H}] = 0.
\l{1.7}
\eeq
В более сложных случаях способы учета  связей  следует  модифицировать
[8,9].

Можно показать,  что при  определенных  предположениях  способы  учета
связей эквивалентны  друг  другу.  Используя  квазиклассические методы
Маслова [10,11], удается перейти от  квантовой  теории  систем  со  связями  к
классической [12].

В работе  строятся  основные  математические  объекты  для  системы со
связями, исследуются их свойства  в  простейшем  случае,  когда  связи
$\hat{\Lambda}_a$ являются линейными комбинациями операторов координат
и импульсов \r{1.5}
\beq
\hat{\Lambda}_a = \sum_i
({\Cal P}_i^{(a)} \hat{q}_i -
{\Cal Q}_i^{(a)} \hat{p}_i), \qquad a = \overline{1,k},
\l{1.8}
\eeq
тогда как   гамильтониан   $\hat{H}$  является  квадратичным  по  этим
операторам:
\beq
\hat{H} =
\frac{1}{2} \sum_{ij}
[\hat{p}_i H_{P_iP_j} \hat{p}_j +
\hat{q}_i H_{Q_iP_j} \hat{p}_j +
\hat{p}_i H_{P_iQ_j} \hat{q}_j +
\hat{q}_i H_{Q_iQ_j} \hat{q}_j]
\l{1.9}
\eeq
Как вытекает  из теории комплексного ростка Маслова в точке [11] для систем
со связями [12],  именно  к  такому  случаю  и  сводится  главный   порядок
квазиклассического разложения.  Рассматриваемые  системы  с  линейными
связями и   квадратичными   гамильтонианами   возникают   также    при
использовании теории  лагранжевых  многообразий  с комплексным ростком
для систем без связей,  в частности в задачах с  вырожденными  точками
покоя и интегралами движения [11,13].

\section{Гильбертово пространство  состояний  системы со связями и его
свойства}

В настоящем  разделе  строится  гильбертово   пространство   состояний
системы со   связями.   В   пункте   2.1   математически  определяется
гильбертово пространство   функций    с    дельтаобразным    скалярным
произведением \r{5}   и   исследуются   его  свойства.  В  пункте  2.2
определяется гильбертово     пространство     обобщенных      функций,
удовлетворяющих свойству    \r{1.6},    и   строится   изоморфизм   этих
гильбертовых пространств.  В пункте 2.3 доказываются леммы и  теорема,
сформулированные в пп. 2.1 и 2.2.

\subsection{Пространство $\Cal H$ и его свойства}

Пусть $\Cal  M  = {\Bbb R}^{2n} = \{(P,Q) | P,Q \in {\Bbb R}^n \}$ ---
$2n$-мерное пространство,  в котором введена кососимметричная  2-форма
("кососкалярное произведение") вида
\beq
\left\langle
\left( \matrix P' \\  Q' \endmatrix \right),
\left( \matrix P'{}' \\  Q'{}' \endmatrix \right)
\right\rangle =
\sum_{i=1}^n (P_i' Q_i'{}' - P_i'{}' Q_i').
\l{2.1}
\eeq
Будем называть {\it  $k$-мерной  изотропной  плоскостью}  в  $\Cal  M$
всякое $k$-мерное  линейное  подпространство ${\Cal L}_k \subset {\Cal
M}$, любые два вектора которого косоортогональны друг другу:
$$
X',X'{}' \in {\Cal L}_k \Rightarrow <X',X'{}'> = 0.
$$
Предположим, что  на  $k$-мерной  изотропной  плоскости  ${\Cal  L}_k$
выбрана мера   интегрирования   $d\mu(X)$,  инвариантная  относительно
сдвигов.

Сопоставим каждому вектору $X = (P,Q) \in {\Cal M}$ оператор
\beq
\Omega(X) \equiv \Omega(P,Q) = \sum_{i=1}^n (P_i \xi_i -
Q_i \frac{1}{i} \frac{\partial}{\partial \xi_i}),
\l{2.2}
\eeq
действующий в пространстве Шварца $S({\Bbb R}^n)$.

{\bf Лемма 2.1. }
{\it
Справедливы свойства:
\beq
(e^{i\Omega(P,Q)} f)(\xi) = e^{i \sum_{j=1}^n (P_j\xi_j -
\frac{1}{2} P_jQ_j)} f(\xi-Q);
\l{2.3}
\eeq
\beb
e^{i\Omega(X)} e^{i\Omega(X')}   =   e^{i\Omega(X+X')}  e^{\frac{i}{2}
<X,X'>}, \\
\Omega(X) e^{i\Omega(X')} = e^{i\Omega(X')}
(\Omega(X) + <X,X'>)
\l{2.4}
\eeb
}

Доказательства всех лемм приведены в пункте 2.3.

Введем в $S({\Bbb R}^n)$ скалярное произведение следующим образом:
\beq
((f',f)) = \int_{{\Cal L}_k} d\mu(X) (f',e^{i\Omega(X)} f),
\l{2.5}
\eeq
где
\bez
(f',f) = \int_{{\Bbb R}^n} d\xi f^{\prime *} (\xi) f(\xi) -
\eez
обычное скалярное произведение.

Корректность определения \r{2.5} обеспечивается леммой 2.2.

{\bf Лемма  2.2.}
{\it
1. Пусть  $f,f'  \in  S({\Bbb  R}^n)$.  Тогда  функция
$(f',e^{i\Omega(X)}f)$, зависящая  от $X \in {\Cal L}_k$,  принадлежит
пространству Шварца. В частности, интеграл \r{2.5} сходится. 
\newline
2. Скалярное  произведение \r{2.5} непрерывно по $f$ и $f'$ в топологии
пространства Шварца.
}

Обозначим через $S({\Bbb  R}^n,{\Cal  L}_k)$  пространство  Шварца  со
скалярным произведением \r{2.5}.

{\it Замечание.}  С  точностью  до  числового множителя формула \r{2.5}
совпадает с \r{5}.  Действительно,  пусть  $({\Cal  X}^{(1)},...,{\Cal
X}^{(k)})$ (${\Cal X}^{(a)} \equiv ({\Cal P}^{(a)}, {\Cal Q}^{(a)})$)
---  базис  на  ${\Cal  L}_k$.  Тогда на ${\Cal L}_k$ можно
ввести координаты $\alpha_1,...,\alpha_k$, сопоставив каждому элементу
$X \in {\Cal L}_k$ его разложение по базису
\beq
X = \sum_{a=1}^k \alpha_a {\Cal X}^{(a)}.
\l{2.6}
\eeq
Инвариантная мера в рассматриваемой системе координат запишется как
\beq
d\mu(X) = J d\alpha_1 ... d\alpha_k, \qquad J = const.
\l{2.7}
\eeq
Поскольку операторы     $\Omega      ({\Cal      X}^{(a)})      \equiv
\hat{\Lambda}^{(a)}$ совпадают с \r{1.8},  скалярное произведение \r{2.5}
примет вид
\beq
((f',f)) = J \int d\alpha_1 ... d\alpha_k
(f',e^{i\alpha_1 \hat{\Lambda}_1 + ... + i\alpha_k \hat{\Lambda}_k} f).
\l{2.8}
\eeq
Как вытекает из леммы 2.1 и свойства изотропности,
$e^{i\alpha_1 \hat{\Lambda}_1 + ... + i\alpha_k \hat{\Lambda}_k}  =
e^{i\alpha_1 \hat{\Lambda}_1} ... e^{i\alpha_k \hat{\Lambda}_k}$.
Поскольку $\int  d\alpha_a e^{i\alpha_a \hat{\Lambda}_a} = 2\pi \delta
(\hat{\Lambda}_a)$, формула  \r{2.8}  действительно  приводится  к виду
\r{5}.

{\bf Лемма 2.3.  } {\it Скалярное произведение  \r{2.5}  неотрицательно
определено:
\beq
((f,f)) \ge 0.
\l{2.9}
\eeq
}

Оказывается, что неравенство  \r{2.9}  является  нестрогим:  существуют
такие ненулевые   функции  $f$,  что  $((f,f))  =  0$.  Действительно,
рассмотрим скалярное  произведение  в  форме   \r{2.8}.   Воспользуемся
тождеством
\bez
J \int d\alpha_1 ... d\alpha_k \frac{1}{i}
\frac{\partial}{\partial \alpha_a}
(f', e^{i\alpha_1  \hat{\Lambda}_1 + ...  + i\alpha_k \hat{\Lambda}_k}
f) = 0,
\eez
которое можно записать также в виде
\beq
((f',\hat{\Lambda}_a f)) = 0
\l{2.10}
\eeq
Подставляя $f'= \hat{\Lambda}_a f$, получим, что
$((\hat{\Lambda}_a f, \hat{\Lambda}_a f)) = 0$. Из соотношения \r{2.10}
также вытекает, что
\beq
((f',\Omega(X) f)) = 0
\l{2.11}
\eeq
для любого $X \in {\Cal L}_k$.

Введем на    пространстве   $S({\Bbb   R}^n,{\Cal   L}_k)$   отношение
эквивалентности. Положим $f_1 \sim f_2$, если
\bez
((f_1-f_2,f_1-f_2)) = 0.
\eez
Тот факт,  что $\sim$ является отношением эквивалентности,  вытекает из
следующей леммы.

{\bf Лемма 2.4. }
{\it
1. Если $((f,f))=0$, то $((f,g)) = 0$ для любого $g$.
\newline
2. Введенное  отношение $\sim$ удовлетворяет свойствам рефлексивности,
симметричности и транзитивности.
}

Рассмотрим факторпространство
$S({\Bbb R}^n,{\Cal   L}_k)   /   \sim$,
состоящее    из    классов
эквивалентности $[f]$.
Сложение, умножение и скалярное произведение  классов  эквивалентности
определим по формулам
\bez
[f_1] + [f_2] = [f_1+f_2]; \quad \alpha[f]=[\alpha f];
\quad ([f_1],[f_2]) = (f_1,f_2)
\eez
Как вытекает из леммы 2.4, введенные определения корректны, то есть не
зависят от  выбора  представителей  классов  эквивалентности  $[f_1]$,
$[f_2]$.

Рассмотрим пополнение факторпространства
\bez
{\Cal H} = \overline{S({\Bbb R}^n,{\Cal   L}_k)   /   \sim}
\eez
Именно оно и выбирается в качестве гильбертова пространства в одном из
подходов к квантованию систем со связями.

Исследуем свойства некоторых операторов,  действующих  в  пространстве
${\Cal H}$. Рассмотрим оператор вида
\bez
e^{i\Omega(Y)}:
S({\Bbb   R}^n,{\Cal   L}_k)
\to
S({\Bbb   R}^n,{\Cal   L}_k)
\eez

{\bf Лемма 2.5.}
{\it
1. Пусть точка $Y$ фазового пространства
принадлежит косоортогональному дополнению к ${\Cal L}_k$
($Y \in {\Cal L}_k^{<\perp>}$), то есть
$<Y,X>=0$ при всех $X \in {\Cal L}_k$. Тогда
оператор $\Omega(Y)$,  рассматриваемый  на  ${S}({\Bbb R}^n,{\Cal
L}_k)$, сохраняет отношение эквивалентности  и  удовлетворяет  условию
эрмитовости
\beq
((f,\Omega(Y)f')) = ((\Omega(Y)f,f')),
\l{2.12}
\eeq
а оператор
$e^{i\Omega(Y)}$ сохраняет скалярное произведение
$((f',f))$ и отношение эквивалентности,
однозначно продолжается на $\Cal H$ до унитарного оператора
$e^{i\Omega(Y)}$.
\newline
2. Для любой последовательности $Y_n \to 0$  из  ${\Cal  L}_k^{<\perp>}$
последовательность унитарных   операторов   $e^{i\Omega(Y_n)}$  сильно
сходится к единичному.
}

{\it Замечание.}  При  $Y  \notin  {\Cal   L}_k^{<\perp>}$   операторы
$e^{i\Omega(Y)}$ и $\Omega(Y)$ не сохраняют отношение эквивалентности,
поэтому их нельзя  распространять  на  $\Cal  H$.  Действительно,  при
$<Y,X> = \alpha \ne 0$ имеем $[\Omega(Y);\Omega(X)] = - i\alpha \ne 0$; тогда
$\Omega(X) f \sim 0$, $\Omega(X) \Omega(Y) f \sim 0$, но
$\Omega(Y) \Omega(X)  f  -  \Omega(X)  \Omega(Y)  f  =  - i\alpha  f$ не
эквивалентно нулю.

\subsection{Пространство
$\check{\Cal  H}$  и  его  изоморфизм  пространству $\Cal H$}

Конструкция гильбертова  пространства $\Cal H$ достаточно сложна:  при
его построении использованы операции факторизации и пополнения. Удобно
использовать также   и   другую  реализацию  гильбертова  пространства
состояний, которая строится в настоящем пункте.

Сопоставим функции
$f \in S({\Bbb   R}^n,{\Cal   L}_k)$
обобщенную функцию $F$ вида
\beq
F = \int_{{\Cal L}_k} d\mu(X) e^{i\Omega(X)} f \equiv \eta f.
\l{2.13}
\eeq
Корректность данного определения вытекает из леммы 2.2.
Обозначим множество  обобщенных функций вида \r{2.13} через
$\check{S} ({\Bbb   R}^n,{\Cal   L}_k)$.

{\bf Лемма 2.6.}
{\it
Пусть
$f \in S({\Bbb   R}^n,{\Cal   L}_k)$ и
$F = \eta f \in \check{S} ({\Bbb   R}^n,{\Cal   L}_k)$.
Тогда $f \sim 0$ тогда и только тогда, когда
$F=0$.
}

Лемма 2.6 позволяет ввести взаимно однозначное отображение
$\eta_0$ пространств $S({\Bbb   R}^n,{\Cal   L}_k)/\sim$ и
$\check{S}({\Bbb   R}^n,{\Cal   L}_k)$ по формуле
\bez
\eta_0 [f] = \eta f, \quad f \in [f].
\eez

{\bf Лемма 2.7.} {\it
1. Пусть  $\{[f_n]\}$  ---  фундаментальная   последовательность   классов
эквивалентности из
$S({\Bbb   R}^n,{\Cal   L}_k) /\sim $.
Тогда последовательность
$F_n = \eta_0 [f_n] \in
\check{S}({\Bbb   R}^n,{\Cal   L}_k)$
сходится в топологии обобщенных функций.
\newline
2. Если
$\eta_0 [f_n] \to_{n\to\infty} 0$ в смысле обобщенных функций, то
$((f_n,f_n)) \to_{n\to\infty} 0$.
}

Из первого  пункта  леммы   2.7   вытекает,   что   каждому   элементу
$\overline{f} \in {\Cal H}$ однозначно сопоставлена обобщенная функция
$F$, которую обозначим как $F \equiv \overline{\eta} \overline{f}$. Из
второго пункта  леммы  2.7  следует,  что  оператор  $\overline{\eta}$
осуществляет взаимно однозначное отображение  $\Cal  H$  на  некоторое
подпространство в пространстве
$S'({\Bbb R}^n)$. Обозначим это подпространство как
\beq
\check{\Cal H} = \overline{\eta} {\Cal H}.
\l{2.14}
\eeq
Определим скалярное произведение в $\check{\Cal H}$ по формуле
\beq
((\overline{\eta} \overline{f},
\overline{\eta} \overline{f}))^{\vee}
\equiv ((\overline{f},\overline{f})).
\l{2.15}
\eeq
Из взаимной однозначности отображения $\overline{\eta}$ вытекает,  что
данное определение корректно.

Опишем класс обобщенных функций,  принадлежащих $\check{\Cal H}$.  Для
этого введем некоторые вспомогательные понятия.

Будем называть $k$-мерную изотропную  плоскость  ${\Cal  G}_k  \subset
{\Cal M}$  {\it  калибровочной  плоскостью} для изотропной плоскости
${\Cal L}_k$,  если  на  ${\Cal  L}_k  +  {\Cal  G}_k$   кососкалярное
произведение невырождено.

{\bf Лемма 2.8.}
{\it
1. Для каждой изотропной  плоскости ${\Cal  L}_k$  можно  подобрать
калибровочную плоскость ${\Cal G}_k$.
\newline
2. Пусть  ${\Cal G}_k$ --- калибровочная плоскость для ${\Cal L}_k$.
Тогда для любого базиса ${\Cal X}^{(1)}$,..., ${\Cal X}^{(k)}$ на
${\Cal L}_k$ можно подобрать базис
${\Cal Y}^{(1)}$,...,    ${\Cal    Y}^{(k)}$    на    ${\Cal    G}_k$,
удовлетворяющий свойству
\beq
<{\Cal X}^{(a)}, {\Cal Y}^{(b)}> = \delta^{ab}.
\l{2.16}
\eeq
3. Пусть ${\Cal G}_k$ --- калибровочная плоскость для ${\Cal  L}_k$.
Тогда $\Cal M$ раскладывается в прямую сумму
\bez
{\Cal M} = {\Cal L}_k + {\Cal G}_k +
({\Cal L}_k + {\Cal G}_k)^{<\perp>};
\eez
здесь через
$({\Cal L}_k   +  {\Cal  G}_k)^{<\perp>}$  обозначено  косоортогональное
дополнение к ${\Cal L}_k + {\Cal G}_k$.
}

Будем говорить,  что обобщенная функция $\Phi  =  \Phi(x,y)$ ($x  \in
{\Bbb R}^{n_1}$, $y \in {\Bbb R}^{n_2}$) из
$S'({\Bbb R}^{n_1+n_2})$ {\it допускает сужение на поверхность $y=0$},
если для любой функции $\varphi \in S({\Bbb R}^{n_1})$ из пространства
Шварца обобщенная функция
\bez
\Phi_{\varphi_1} (y) = \int dx \varphi_1(x) F(x,y)
\eez
(очевидно принадлежащая  $S'({\Bbb  R}^{n_2})$)  является  непрерывной
функцией $y$ в некоторой окрестности точки $y=0$.

Рассмотрим выражение вида
\beq
R(Y) = (F',e^{i\Omega(Y)} F), \qquad Y \in {\Cal M}
\l{2.17}
\eeq

{\bf Лемма 2.9.} {\it
При $F,F' \in S'({\Bbb R}^n)$ выражение \r{2.17}  определяет  обобщенную
функцию переменной  $Y \in {\Bbb R}^{2n}$ из $S'({\Bbb R}^{2n})$.  Эта
обобщенная функция непрерывно зависит от $F$ и $F'$.
}

Класс $\check{\Cal  H}$  обобщенных  функций можно описать,  используя
следующую теорему.

{\bf Теорема 2.1.}
{\it
Для того чтобы обобщенная функция  $F  \in  S'({\Bbb
R}^n)$ принадлежала классу $\check{\Cal H}$,  необходимо и достаточно,
чтобы вместе выполнялись следующие два условия:

(а)
\bez
\Omega(X) F = 0
\eez
для всех $X \in {\Cal L}_k$;

(б) обобщенная функция
\bez
(F, e^{i\Omega(Y)} F)
\eez
допускает сужение на некоторую калибровочную плоскость ${\Cal G}_k$.
}

Скалярное произведение  в пространстве $\check{\Cal H}$ можно записать
явно (без использования косвенной формулы \r{2.15}) через \r{2.17}.

Пусть ${\Cal L}_k$ --- изотропная плоскость в $\Cal M$,  ${\Cal  G}_k$
--- калибровочная поверхность для нее; на
${\Cal L}_k$ и ${\Cal  G}_k$  заданы  меры  $d\mu(X)$  и  $d\sigma(Y)$
соответственно, инвариантные относительно сдвигов.

{\bf Лемма 2.10.} {\it
1. Для некоторой константы $\Delta$,  зависящей  от  вида  мер  $d\mu$  и
$d\sigma$ и не зависящей от функции $\rho$, справедливо свойство
\beq
\int_{{\Cal L}_k} d\mu(X)
\int_{{\Cal G}_k} d\sigma(Y) \rho(Y)
e^{i<X,Y>} = \rho(0) \times \Delta.
\l{2.18}
\eeq
2. Для любой функции $\rho(Y)$,  $Y \in {\Cal G}_k$,  из  пространства
Шварца, удовлетворяющей  свойству  $\rho(0)  = 1/\Delta$,  справедливо
свойство
\beq
((F,F))^{\vee} = \int_{{\Cal G}_k} d\sigma(Y) \rho(Y)
(F,e^{i\Omega(Y)} F), \qquad F \in \check{\Cal H}.
\l{2.19}
\eeq
}

Рассмотрим примеры использования полученных формул.

Прежде всего, отметим, что в частном случае системы без связей теорема
2.1 дает новый критерий принадлежности  обобщенной  функции  $F  \in
S'({\Bbb R}^n)$ пространству $L^2$.

{\bf Теорема 2.2}.
{\it
Для того  чтобы обобщенная функция $F \in S'({\Bbb R}^n)$ принадлежала
пространству $L^2$, необходимо и достаточно, чтобы обобщенная функция
\bez
(F, e^{i\sum_j (P_j\hat{q}_j - Q_j\hat{p}_j) } F) = \rho(P,Q)
\eez
из $S'({\Bbb R}^{2n})$ была непрерывна в некоторой  окрестности  точки
$(P,Q) = 0$.
}

Рассмотрим теперь простейшие примеры систем со связями.

{\it Пример 1.} Пусть $n=1,k=1$ (система с одной  степенью  свободы  и
одной связью).  Пусть  изотропная  плоскость  ${\Cal L}_1$ --- прямая,
натянутая на вектор $(P=0,Q=1)$. Тогда гильбертово пространство ${\Cal
H}$ строится   как  пополнение  факторпространства  $S({\Bbb  R};{\Cal
L}_1)$ со скалярным произведением
\beq
((f,f)) = 2\pi (f,\delta(\hat{p}_1) f) = |\int d\xi f(\xi)|^2.
\l{2.20}
\eeq
Очевидно, что  две  функции  $f$ эквивалентны,  если равны
интегралы $\int d\xi f(\xi)$; тем самым каждому классу эквивалентности
однозначно сопоставлено   число  ---  интеграл  $\int  d\xi  f(\xi)  =
\overline{f}$. Тем   самым   пространство   ${\Cal   H}$   оказывается
одномерным.

Построим теперь  гильбертово  пространство $\check{\Cal H}$.  Согласно
теореме 2.1,  оно состоит из обобщенных функций $F \in S'({\Bbb  R})$,
удовлетворяющих двум условиям. Первое из них записывается как
$\hat{p} F  =  0$,  или  $\frac{\partial  F}{\partial   \xi}   =   0$.
Следовательно, $F=const$.

Чтобы исследовать    второе   условие,   отметим,   что   в   качестве
калибровочной поверхности    можно    выбрать     любое     одномерное
подпространство, натянутое на вектор $(P,Q)$ при $P\ne 0$.  Натянем ее
на вектор $(P=1,Q=0)$. Тогда при $Y = (\beta,0) \in {\Cal G}_1$ получим
\bez
(F, e^{i\Omega(Y)} F) = (F, e^{i\beta \hat{q}} F) =
2\pi |F|^2 \delta(\beta).
\eez
Видно, что   для   функции  $F=const$  скалярное  произведение  \r{2.17}
действительно допускает  сужение  на  калибровочную  поверхность.  Тем
самым пространство $\check{\Cal H}$ также оказывается одномерным.

Связь функций $F \in \check{\Cal H}$ и $f \in {\Cal H}$ выглядит так:
\bez
F = 2\pi \delta(\hat{p}) f,
\eez
или
\bez
F(\xi) = \int d\xi f(\xi+\alpha) = \int d\alpha f(\alpha).
\eez
Из формул \r{2.18} и \r{2.19} находим, что $((F,F))^{\vee} = |F|^2$, что
согласуется с \r{2.20}.

{\it Пример  2.}  Аналогичным  образом  рассматривается  и  пример   с
$n=1$, $k=1$,  когда  подпространство  ${\Cal  L}_1$  является прямой,
натянутой на вектор $(P=1,Q=0)$. Тогда
\bez
((f,f)) = 2\pi (f,\delta(\hat{q}) f) = 2\pi |f(0)|^2.
\eez
Следовательно, две  функции  $f$  эквивалентны,  если  совпадают   при
$\xi=0$, и  каждому  классу  эквивалентности  сопоставлено  число  ---
значение этой функции в точке $\xi=0$ (для  удобства  умножим  его  на
$\sqrt{2\pi}$): $\overline{f} = \sqrt{2\pi} f(0)$.
Тем самым $\check{\Cal H}$ одномерно.

Первое условие теоремы 2.1 на функцию $F\in \check{\Cal H}$ имеет  вид
$\hat{q} F  =  0$,  что  означает  $\xi  F(\xi)  =  0$,  или $F(\xi) =
\sqrt{2\pi} F \delta(\xi)$.  Нетрудно показать,  что и второе  условие
теоремы 2.1   для   функции   $F$   данного   вида   также  выполнено.
Следовательно, пространство $\check{\Cal H}$ также одномерно.

Изоморфизм пространств $\Cal H$ и $\check{\Cal H}$ строится с  помощью
оператора $\eta$
\bez
\eta f(\xi)  =  2\pi  \delta(\xi)  f(\xi)  =  \sqrt{2\pi} \overline{f}
\delta(\xi).
\eez
Как нетрудно показать, $((F,F))^{\vee} = |F|^2$.

Тривиальные примеры 1 и 2 можно использовать как основу для построения
более сложных, нетривиальных примеров.

{\it Пример 3.}
Пусть $n=2,k=1$,  ${\Cal L}_1$ --- прямая,  натянутая на вектор $(P=0,
Q_1=1, Q_2=0)$.  Тогда $\Cal H$ --- это пополнение  факторпространства
$S'({\Bbb R}^2)$ со скалярным произведением
\bez
((f,f)) = 2\pi (f, \delta(\hat{p}_1) f) =
\int d\xi_2 |\int d\xi_1 f(\xi_1,\xi_2)|^2.
\eez
При этом две функции $f$ эквивалентны, если равны интегралы
$F(\xi_2) = \int d\xi_1 f(\xi_1,\xi_2)$.  Тем самым пространство $\Cal
H$ эквивалентно пространству $L^2({\Bbb R})$.

Аналогично, первое условие теоремы 2.1 записывается как
$\hat{p}_1 F = 0$, или $F = F(\xi_2)$. Нетрудно показать, что
$((F,F))^{\vee} = \int d\xi_2 |F(\xi_2)|^2$, а второе условие
теоремы 2.1 эквивалентно принадлежности $F$ пространству $L^2$.

\subsection{Доказательства лемм и теоремы}

Докажем утверждения, сформулированные в настоящем разделе.

{\it Доказательство  леммы  2.1.}  Экспоненту  от  оператора $e^{i\tau
\Omega(P,Q)}$ можно определить  как  оператор,  переводящий  начальное
условие $f_0(\xi)$ задачи Коши для эволюционного уравнения
\bez
- i \frac{\partial f(\tau,\xi)}{\partial \tau} =
\Omega(P,Q) f(\tau,\xi) =
\sum_{j=1}^n (P_j   \xi_j  -  Q_j  \frac{1}{i}\frac{\partial}{\partial
\xi_j})
f(\tau,\xi)
\eez
в решение этой задачи $f(\tau,\cdot)$ "в момент времени" $\tau$. Явной
подстановкой находим, что
\bez
f(\tau,\xi) = e^{i
\sum_{j=1}^n (\tau P_j   \xi_j  - \frac{\tau^2}{2} P_jQ_j)
} f_0(\xi - Q\tau),
\eez
что доказывает   формулу   \r{2.3}.    Формулы    \r{2.4}    получаются
комбинированием формул \r{2.3}.

{\it Доказательство леммы 2.2.} Используя формулу \r{2.3}, получим
\beq
(f', e^{i\Omega(P,Q)} f) =
\int d\xi f^{\prime *}(\xi)
e^{i
\sum_{j=1}^n (P_j   \xi_j  - \frac{1}{2} P_jQ_j)
} f(\xi - Q).
\l{2.21}
\eeq
Разлагая функцию $f(\xi)$ в интеграл Фурье
\bez
f(\xi) = \int d\kappa e^{i\kappa \xi} \tilde{f}(\kappa),
\eez
приводим выражение \r{2.21}  к  виду  преобразования  Фурье  функции  из
пространства Шварца
\bez
(f', e^{i\Omega(P,Q)} f) =
\int d\xi d\kappa f^{\prime *}(\xi)
e^{i\kappa \xi} \tilde{f}(\kappa)
e^{i
\sum_{j=1}^n (P_j   \xi_j  - \frac{1}{2} P_jQ_j - \kappa_j Q_j)
},
\eez
которая также принадлежит пространству Шварца и непрерывно зависит  от
$f$ и  $f'$.  Такими же свойствами обладает и интеграл от этой функции
по любой плоскости.

{\it Доказательство леммы 2.3.}  Введем  "регуляризованное"  скалярное
произведение вида
\beq
((f',f))_{\varepsilon} = \int_{{\Cal L}_k} d\mu(X)
e^{-{\varepsilon} M(X,X)} (f', e^{i\Omega(X)} f),
\l{2.22}
\eeq
где $M(X,X)$  ---  произвольная положительно определенная квадратичная
форма. Из теоремы Лебега и принадлежности функции
$(f', e^{i\Omega(X)} f)$ пространству Шварца вытекает, что
\beq
((f',f))_{\varepsilon} \to_{{\varepsilon} \to 0} ((f',f)).
\l{2.23}
\eeq
Проверим теперь неотрицательную определенность скалярного произведения
\r{2.22}, равносильную неотрицательной определенности оператора
\beq
\eta_{\varepsilon} =
\int_{{\Cal L}_k} d\mu(X)
e^{-{\varepsilon} M(X,X)}  e^{i\Omega(X)},
\l{2.24}
\eeq
ограниченного ввиду оценки
\bez
||\eta_{\varepsilon}|| \le
\int_{{\Cal L}_k} d\mu(X)
e^{-{\varepsilon} M(X,X)}.
\eez
Представим оператор  $\eta_{\varepsilon}$  в  виде  квадрата  эрмитова
оператора $\mu_{\varepsilon}$, который будем искать в виде
\bez
\mu_{\varepsilon} = C_1
\int_{{\Cal L}_k} d\mu(X)
e^{-2{\varepsilon} M(X,X)} e^{i\Omega(X)}.
\eez
Учитывая, что $<X,X'> = 0$, по лемме 2.1 получим
\bez
\mu_{\varepsilon}^+ \mu_{\varepsilon} = C_1^2
\int_{{\Cal L}_k} d\mu(X')
\int_{{\Cal L}_k} d\mu(X'{}')
e^{-2{\varepsilon} M(X',X')}
e^{-2{\varepsilon} M(X'{}',X'{}')}
e^{i\Omega(X' + X'{}')}.
\eez
Перейдем к переменным интегрирования $X'$ и $X = X' + X'{}'$; получим,
что
\bez
\mu_{\varepsilon}^+ \mu_{\varepsilon} =
\int_{{\Cal L}_k} d\mu(X) e^{i\Omega(X)} \rho(X),
\eez
где числовая функция $\rho(X)$ имеет вид
\bey
\rho(X) = C_1^2
\int_{{\Cal L}_k} d\mu(X') e^{-2{\varepsilon}
[M(X',X') + M(X-X',X-X')]} =\\
e^{-{\varepsilon} M(X.X)}
C_1^2 \int_{{\Cal L}_k} d\mu(X') e^{-4{\varepsilon}
M(X' - \frac{X}{2}, X' - \frac{X}{2})}.
\eey
При некотором  $C_1$  получим  $\rho(X) = e^{- {\varepsilon} M(X,X)}$,
так что
\bez
\eta_{\varepsilon} = \mu_{\varepsilon}^+ \mu_{\varepsilon}.
\eez
Следовательно,
\bez
((f,f))_{\varepsilon} = (f,\eta_{\varepsilon} f) =  (\mu_{\varepsilon}
f, \mu_{\varepsilon} f) \ge 0.
\eez
Используя \r{2.23}, получаем утверждение леммы.

{\it Доказательство  леммы 2.4.} Первое утверждение леммы 2.4 вытекает
из неравенства  Коши-Буняковского,  которое  для  вырожденного  случая
доказывается так  же,  как  и  для  невырожденного.  Рефлексивность  и
симметричность отношения $\sim$ очевидны.  Транзитивность вытекает  из
доказанного первого пункта.

{\it Доказательство  леммы  2.5.}  Свойства
оператора $\Omega(Y)$ вытекают из определения этого оператора.
Сохранение  скалярного произведения
вытекает из  перестановочности  оператора  $e^{i\Omega(Y)}$  со  всеми
операторами $e^{i\Omega(X)}$ (в силу леммы 2.1 и свойства $<Y,X>=0$),
а значит,  и  с   оператором   $\eta_{\varepsilon}$.   Из   сохранения
скалярного произведения  вытекает сохранение отношения эквивалентности
и однозначная продолжаемость до изометрического оператора в $\Cal  H$.
Из свойства    $(e^{i\Omega(Y)})^{-1}   =   e^{i\Omega(Y)}$   вытекает
обратимость и унитарность.

Чтобы доказать  второе  утверждение  леммы,  рассмотрим   произвольную
функцию $f \in S({\Bbb R}^n)$. Докажем, что
\beq
||(e^{i\Omega(Y_n)} - 1) f||_{\Cal H} \to 0
\l{2.25}
\eeq
Имеем:
\bez
||(e^{i\Omega(Y_n)} - 1) f||_{\Cal H}  =
||i \int_0^1 d\tau e^{i\tau \Omega(Y_n)} \Omega(Y_n) f||_{\Cal H}  =
||\Omega(Y_n)f||_{\Cal H}.
\eez
Поскольку $\Omega(Y_n)f$ -- линейная комбинация функций $\xi_i f(\xi)$
и $\frac{\partial f}{\partial \xi_i}$ с коэффициентами, стремящимися к
нулю, $||\Omega(Y_n)f||_{\Cal  H}  \to  0$.  Следовательно,   свойство
\r{2.25} проверено    для    $f   \in   S({\Bbb   R}^n)$.   По   теореме
Банаха-Штейнгауза оно может быть продолжено на $\Cal H$.

{\it Доказательство леммы 2.6.} Пусть $f\sim 0$.  Тогда при любом  $f'
\in S({\Bbb R}^n)$
\beq
(f',F) = ((f',f)) = 0,
\l{2.26}
\eeq
как вытекает из неравенства Коши-Буняковского. Следовательно, $F=0$.

Обратно, пусть  $F=0$.  Тогда  для  любой  $f'  \in   S({\Bbb   R}^n)$
справедливо свойство  \r{2.26}.  В частности,  оно выполнено при $f'=f$,
так что $f\sim 0$.

{\it Доказательство леммы 2.7.} Если $\{ [f_n] \}$ --- фундаментальная
последовательность классов эквивалентности из
$S({\Bbb R}^n,{\Cal L}_k)/\sim$,
то для любого ${\varepsilon} < 0$
\bez
||f_n - f_{n'}||_{\Cal H} < {\varepsilon},
\eez
начиная с  некоторых  $n,n'  > N({\varepsilon})$.  Следовательно,  для
любого $f' \in S({\Bbb R}^n)$ по неравенству Коши-Буняковского
\bez
|((f_n-f_{n'},f')|| \le ||f'||_{\Cal H} \sqrt{\varepsilon},
\eez
так что числовая последовательность
\bez
((f',f_n)) = (f', \eta f_n) = (f',F_n)
\eez
фундаментальна и имеет предел.  Следовательно,  последовательность $\{
F_n\}$ сходится в топологии обобщенных функций.

Докажем второе   утверждение   леммы.   Из   свойства   $\eta_0  [f_n]
\to_{n\to\infty} 0$ вытекает, что
\bez
((f',f_n) \to_{n\to \infty} 0
\eez
для любой  $f'  \in  S({\Bbb   R}^n)$.   Докажем,   что   $((f_n,f_n))
\to_{n\to\infty} 0$.   Предположим   противное;  тогда  для  некоторой
подпоследовательности $f_{n_k}$ справедливы свойства
\bez
((f_{n_k},f_{n_k})) \ge {\varepsilon} > 0;
\qquad
((f',f_{n_k})) \to_{k\to\infty} 0.
\eez
Рассмотрим произвольное $l$. Выберем такое $m=n_k > l$, что
\bez
|((f_l,f_m))| < {\varepsilon}/4.
\eez
Тогда
\bez
|((f_l - f_m,f_l-f_m))| \ge {\varepsilon}/2.
\eez
Полученное противоречие с  фундаментальностью  последовательности  $\{
[f_n]\}$ доказывает лемму.

{\it Доказательство леммы 2.8.}
Пусть ${\Cal L}_k$ --- изотропная плоскость и
$({\Cal X}^{(1)},...,{\Cal X}^{(k)})$ --- базис на ней; при этом
\beq
<{\Cal X}^{(a)}, {\Cal X}^{(b)}>  = 0, \quad a,b = \overline{1,k}
\l{2.27}
\eeq
Продолжим этот базис до базиса
$({\Cal X}^{(1)},...,{\Cal   X}^{(2n)})$   на   $\Cal  M$.  Рассмотрим
линейные функционалы $\omega_j$,  $j=\overline{1,2n}$,  на  $\Cal  M$,
определенные на базисных векторах как
\bez
\omega_j ({\Cal X}^{(i)}) = \delta_{ij}.
\eez
В силу  невыррожденности  кососкалярного  произведения  любой линейный
функционал на $\Cal M$ может быть представлен как
\bez
\omega_j(X) = <X, \tilde{\Cal Y}^{(j)}>
\eez
для некоторого $\tilde{\Cal Y}^{(j)}  \in  {\Cal  M}$.  Следовательно,
выбраны линейно независимые векторы
$\tilde{\Cal Y}^{(1)},...,\tilde{\Cal    Y}^{(k)}$,    удовлетворяющие
условиям
\bez
<{\Cal X}^{(a)},    \tilde{\Cal   Y}^{(b)}>   =   \delta_{ab},   \quad
a,b=\overline{1,k}.
\eez
Рассмотрим теперь векторы
\beq
{\Cal Y}^{(a)} = \tilde{\Cal Y}^{(a)} -
\frac{1}{2} \sum_{c=1}^k
<\tilde{\Cal Y}^{(a)},\tilde{\Cal Y}^{(c)}> {\Cal X}^{(c)}.
\l{2.28}
\eeq
Учитывая \r{2.27}, непосредственным вычислением получаем, что
\beq
<{\Cal Y}^{(a)},{\Cal Y}^{(b)}> = 0;
\quad
<{\Cal X}^{(a)},{\Cal Y}^{(b)}> = \delta_{ab}; \quad
a,b = \overline{1,k}.
\l{2.29}
\eeq
Поскольку $<{\Cal  X}^{(a)},\sum_b  {\Cal  Y}^{(b)}>  = \alpha_a$,  из
свойства $\sum_b \alpha_b {\Cal  Y}^{(b)}>  =  0$  вытекает,  что  все
$\alpha_b=0$. Следовательно, векторы
${\Cal Y}^{(1)},...,{\Cal Y}^{(k)}>$ линейно независимы.  Их  линейная
оболочка ${\Cal   G}_k$  удовлетворяет  всем  свойствам  калибровочной
плоскости. Первое утверждение леммы доказано.

Пусть ${\Cal G}_k$ --- произвольная калибровочная плоскость,
$({\Cal X}^{(1)},...,{\Cal X}^{(k)})$ --- базис на ${\Cal L}_k$,
$({\Cal Y}^{(1)\prime},...,{\Cal Y}^{(k)\prime})$ --- базис на  ${\Cal
G}_k$. Ввиду  условия  невырожденности  кососкалярного произведения на
${\Cal L}_k + {\Cal G}_k$ матрица
\bez
<({\Cal X}^{(a)},{\Cal Y}^{(b)\prime}) = M_{ab}
\eez
также невырождена. Тогда базис
\bez
{\Cal Y}^{(a)} = (M^{-1})_{ca} {\Cal Y}^{(c)\prime}
\eez
удовлетворяет всем требуемым свойствам \r{2.16}.

Докажем третье утверждение леммы. Пусть
$({\Cal X}^{(1)},...,{\Cal X}^{(k)})$
и
$({\Cal Y}^{(1\prime)},...,{\Cal Y}^{(k\prime)})$
--- введенные базисы на ${\Cal L}_k$ и ${\Cal G}_k$.  Любой вектор  $Y
\in {\Cal M}$ можно представить в виде суммы
\bez
Y =
- \sum_{a=1}^k
<{\Cal Y}^{(a)},Y> {\Cal X}^{(a)}
+ \sum_{b=1}^k
<{\Cal X}^{(b)},Y> {\Cal Y}^{(b)} + Z,
\eez
первое слагаемое  которой принадлежит ${\Cal L}_k$,  второе --- ${\Cal
G}_k$, третье --- $({\Cal L}_k + {\Cal G}_k)^{< \perp >}$, поскольку
\bez
<{\Cal X}^{(b)},Z> = 0;
\qquad
<{\Cal Y}^{(a)},Z> = 0.
\eez
Далее, предположим, что
\beq
\sum_{a=1}^k \alpha_a {\Cal X}^{(a)}
+ \sum_{b=1}^k \beta_b {\Cal Y}^{(b)} + Z  = 0;
\quad
Z \in ({\Cal L}_k + {\Cal G}_k)^{< \perp >}.
\l{2.30}
\eeq
Беря косоортогональное  произведение   равенства   \r{2.30}   с   ${\Cal
X}^{(a)}$ и ${\Cal Y}^{(b)}$,  находим, что $\alpha_a=0$, $\beta_b=0$;
следовательно, $Z=0$. Лемма полностью доказана.

{\it Доказательство леммы 2.9.}
Для любой функции $\chi \in S({\Bbb R}^{2n})$ рассмотрим интеграл
\bey
R[\chi] = \int dPdQ \chi(P,Q) (F', e^{i\Omega(P,Q)} F) = \\
\int dPdQ d\xi F^{\prime *}(\xi)
e^{i \sum_{j=1}^n
(P_j\xi_j - \frac{1}{2} P_jQ_j) }
F(\xi-Q) \chi(P,Q).
\eey
Переходя к переменным интегрирования $P,\xi,\eta=\xi-Q$, получаем
\beq
R[\chi] = \int d\xi d\eta F^{\prime *}(\xi) F(\eta)
\int dP e^{i\sum_{j=1}^n P_j \frac{\xi_j+\eta_j}{2}} \chi(P,\xi-\eta).
\l{2.31}
\eeq
Как известно,  прямое  произведение  обобщенных   функций   $F^{\prime
*}(\xi) F(\eta)$  также  является  обобщенной  функцией  из  $S'({\Bbb
R}^{2n})$, непрерывно зависящей от $F$ и $F'$. Функция
\bez
\int dP   e^{i   \sum_{j=1}^n   P_j   \frac{\xi_j   +   \eta_j}{2}   }
\chi(P,\xi-\eta)
\eez
принадлежит классу  $S({\Bbb R}^{2n})$ и непрерывно зависит от $\chi$.
Следовательно, выражение  $R[\chi]$  определяет  линейный  непрерывный
функционал в $S({\Bbb R}^{2n})$, непрерывно зависящий от $F$ и $F'$.

{\it Доказательство теоремы 2.1.}
Докажем необходимость.  Пусть $F \in \check{\Cal H}$.  Установим,  что
$\Omega(X) F = 0$ для всех $X\in {\Cal L}_k$, а обобщенная функция
$(F,e^{i\Omega(Y)}F)$ допускает   сужение   на   любую   калибровочную
плоскость. Доказательство разобьем на несколько лемм.

{\bf Лемма 2.11. } {\it
Пусть $F \in \check{\Cal H}$. Тогда $\Omega(X)F=0$ для
всех $X \in {\Cal L}_k$.
}

{\bf Доказательство.} Пусть $F=\eta f$,  где
$f\in S({\Bbb  R}^n,{\Cal L}_k)$. Тогда для любой функции $f'$ получим
\bez
(f',\Omega(X)f) = (\Omega(X)f',F) = ((\Omega(X)f',f)) = 0
\eez
в силу определения обобщенной производной и свойства \r{2.11}.
Следовательно, $\Omega(X) F = 0$.

Далее, пусть  $F  \in  \check{\Cal  H}$.   Это   означает,   что   $F=
\lim_{n\to\infty} F_n$,   где  $F_n  \in  \check{S}({\Bbb  R}^n,{\Cal
L}_k)$ и $\Omega(X)  F_n  =  0$  по  доказанному.  Поскольку  операции
умножения на  $\xi_i$  и  дифференцирования  $\frac{\partial}{\partial
\xi_i}$ непрерывны в топологии обобщенных функций,  $\Omega(X) F =  0$
для любого $F \in \check{\Cal H}$. Лемма 2.11 доказана.

{\bf Лемма 2.12.} {\it
Пусть $F  \in  \check{S}({\Bbb  R}^n,{\Cal  L}_k)$.  Тогда   обобщенная
функция $(F,   e^{i\Omega(Y)}   F)$   допускает   сужение   на   любую
калибровочную плоскость ${\Cal G}_k$.
}

{\bf Доказательство.}   Пусть   ${\Cal    G}_k$    ---    произвольная
калибровочная плоскость.  Введем  в  ${\Cal  M} \subset {\Cal R}^{2n}$
координаты следующим образом. Пусть
${\Cal X}^{(1)},...,{\Cal X}^{(k)}$ --- базис на ${\Cal L}_k$,
${\Cal Y}^{(1)},...,{\Cal Y}^{(k)}$ --- базис на ${\Cal G}_k$,
удовлетворяющий свойству \r{2.29},
${\Cal Z}^{(1)},...,{\Cal Z}^{(2n-2k)}$ ---
произвольный базис на $({\Cal L}_k + {\Cal G}_k)^{<\perp>}$.
Сопоставим $Y \in {\Cal M}$ разложение
\beq
Y =
\sum_{a=1}^k \alpha_a {\Cal X}^{(a)} + \sum_{b=1}^k \beta_b {\Cal Y}^{(b)}
+ \sum_{\lambda} \gamma_{\lambda} {\Cal Z}^{(\lambda)}.
\l{2.32}
\eeq
Назовем $(\alpha,\beta,\gamma)$  координатами  точки  $Y\in {\Cal M}$.
Утверждение леммы означает, что функция
\beq
(F,
e^{i
\sum_{a=1}^k \alpha_a \Omega({\Cal X}^{(a)})
+ i \sum_{b=1}^k \beta_b \Omega({\Cal Y}^{(b)})
+ i \sum_{\lambda} \gamma_{\lambda} \Omega({\Cal Z}^{(\lambda)})
}
F)
\l{2.33}
\eeq
может рассматриваться  как обобщенная функция $\beta$,  непрерывная по
$\alpha$ и $\gamma$ в окрестности $(\alpha=0,\gamma=0)$.  Покажем, что
это так и что для любой $\chi \in S({\Bbb R}^n)$ функция
\beq
\rho(\alpha,\gamma) = \int d\beta \chi(\beta)
(F,
e^{i
\sum_{a=1}^k \alpha_a \Omega({\Cal X}^{(a)})
+ i\sum_{b=1}^k \beta_b \Omega({\Cal Y}^{(b)})
+ i \sum_{\lambda} \gamma_{\lambda} \Omega({\Cal Z}^{(\lambda)})
}
F)
\l{2.34}
\eeq
является непрерывной    функцией    $\alpha,\gamma$.    Проверим   это
утверждение.

Пусть
\bez
F=\eta f = \int d\alpha
e^{i
\sum_{a=1}^k \alpha_a \Omega({\Cal X}^{(a)})
}
f, \qquad
f \in S({\Bbb R}^n,{\Cal L}_k).
\eez
Тогда для интеграла \r{2.34} получим
\bey
\rho(\alpha,\gamma) =
J^2 \int d\alpha' d\alpha'{}' d\beta
(f,
e^{i
\sum_{a=1}^k \alpha_a' \Omega({\Cal X}^{(a)})
}
\\
e^{i
\sum_{a=1}^k \alpha_a \Omega({\Cal X}^{(a)})
+ i\sum_{b=1}^k \beta_b \Omega({\Cal Y}^{(b)})
+ i\sum_{\lambda} \gamma_{\lambda} \Omega({\Cal Z}^{(\lambda)})
}
e^{i
\sum_{a=1}^k \alpha_a'{}' \Omega({\Cal X}^{(a)})
}
f).
\eey
Используя второе утверждение леммы 2.1,  приведем данное соотношение к
виду
\bey
\rho(\alpha,\gamma) =
J^2 \int d\alpha' d\alpha'{}' d\beta
\chi(\beta)
e^{\frac{i}{2} \sum_{a=1}^k (\alpha_a' - \alpha_a'{}') \beta_a}
\\ \times (f,
e^{i
\sum_{a=1}^k (\alpha_a + \alpha_a' + \alpha_a'{}')
\Omega({\Cal X}^{(a)})
+ \sum_{b=1}^k \beta_b \Omega({\Cal Y}^{(b)})
+ \sum_{\lambda} \gamma_{\lambda} \Omega({\Cal Z}^{(\lambda)})
}
f).
\eey
Перейдем от переменных интегрирования $\alpha'$ и $\alpha'{}'$ к
\bez
\mu_a = \alpha_a + \alpha_a' + \alpha_a'{}';
\qquad
\kappa_a = \frac{\alpha_a' - \alpha_a'{}'}{2}.
\eez
Якобиан такой замены переменных равен единице; отсюда
\bey
\rho(\alpha,\gamma) = J^2 \int d\mu d\kappa d\beta
\chi(\beta) e^{i\sum_{a=1}^k \kappa_a \beta_a} \\ \times
e^{i
\sum_{a=1}^k \mu_a \Omega({\Cal X}^{(a)})
+ i\sum_{b=1}^k \beta_b \Omega({\Cal Y}^{(b)})
+ i\sum_{\lambda} \gamma_{\lambda} \Omega({\Cal Z}^{(\lambda)})
}
\eey
Используя известную формулу
\bez
\int d\kappa   e^{i\sum_{a=1}^k   \kappa_a\beta_a}   =   \prod_a  2\pi
\delta(\beta_a),
\eez
находим
\bez
\rho(\alpha,\gamma) = J^2 (2\pi)^k \chi(0) \int d\mu
(f,
e^{i
\sum_{a=1}^k \mu_a \Omega({\Cal X}^{(a)})}
e^{i
\sum_{\lambda} \gamma_{\lambda} \Omega({\Cal Z}^{(\lambda)})
} f).
\eez
Следовательно,
\bey
(F,
e^{i
\sum_{a=1}^k \alpha_a \Omega({\Cal X}^{(a)})
+ \sum_{b=1}^k \beta_b \Omega({\Cal Y}^{(b)})
+ \sum_{\lambda} \gamma_{\lambda} \Omega({\Cal Z}^{(\lambda)})
}
F)  \\
= J (2\pi)^k \prod_b
\delta(\beta_b)
((f,
e^{i
\sum_{\lambda} \gamma_{\lambda} \Omega({\Cal Z}^{(\lambda)})
} f)).
\eey
При этом функция
$((f, e^{i
\sum_{\lambda} \gamma_{\lambda} \Omega({\Cal Z}^{(\lambda)})
} f))$  непрерывно зависит от $\gamma$,  что вытекает хотя бы из леммы
2.5.

Таким образом,   обобщенная  функция
\r{2.33}  действительно  допускает сужение на калибровочную плоскость.
Лемма доказана.

{\bf Лемма 2.13.}
{\it Пусть  обобщенная  функция  $F$  принадлежит  классу
$\check{\Cal H}$. Тогда   обобщенная
функция $(F,   e^{i\Omega(Y)}   F)$   допускает   сужение   на   любую
калибровочную плоскость ${\Cal G}_k$.
}

{\bf Доказательство.} Условие $F\in \check{\Cal H}$ означает,  что $F=
\lim_{n\to\infty} F_n$,  где  $F_n  =  \eta  f_n$,  и  $\{f_n\}$   ---
последовательность функций  из  пространства  Шварца,  удовлетворяющая
свойству
$((f_n - f_m,f_n - f_m)) \to_{n,m\to\infty} 0$.
При этом
\beb
(F,
e^{i
\sum_{a=1}^k \alpha_a \Omega({\Cal X}^{(a)})
+ \sum_{b=1}^k \beta_b \Omega({\Cal Y}^{(b)})
+ \sum_{\lambda} \gamma_{\lambda} \Omega({\Cal Z}^{(\lambda)})
}
F) = \\
\lim_{n\to\infty}
(F_n,
e^{i
\sum_{a=1}^k \alpha_a \Omega({\Cal X}^{(a)})
+ \sum_{b=1}^k \beta_b \Omega({\Cal Y}^{(b)})
+ \sum_{\lambda} \gamma_{\lambda} \Omega({\Cal Z}^{(\lambda)})
}
F_n) = \\
\lim_{n\to\infty}
J (2\pi)^k \prod_b
\delta(\beta_b)
((f_n,
e^{i
\sum_{\lambda} \gamma_{\lambda} \Omega({\Cal Z}^{(\lambda)})
} f_n)) = \\
J (2\pi)^k \prod_b
\delta(\beta_b)
((f,
e^{i
\sum_{\lambda} \gamma_{\lambda} \Omega({\Cal Z}^{(\lambda)})
} \overline{f})).
\l{2.35}
\eeb
Отсюда следует утверждение леммы.

Докажем достаточность в теореме 2.1. Пусть
\beq
\Omega(X) F = 0, \quad X \in {\Cal L}_k,
\l{2.36}
\eeq
а интеграл
\beq
\rho(\alpha,\beta,\gamma) =
(F,
e^{i
\sum_{a=1}^k \alpha_a \Omega({\Cal X}^{(a)})
+ \sum_{b=1}^k \beta_b \Omega({\Cal Y}^{(b)})
+ \sum_{\lambda} \gamma_{\lambda} \Omega({\Cal Z}^{(\lambda)})
}
F)
\l{2.37}
\eeq
является обобщенной функцией $\beta$, непрерывной по $\alpha,\gamma$ в
окрестности точки $\alpha=0,\gamma=0$.
Требуется построить  фундаментальную   последовательность   $f_n   \in
S({\Bbb R}^n,{\Cal  L}_k)$,  для  которой  $\eta f_n$ сходится к $F$ в
обобщенном смысле:
\beq
((f_n-f_m,f_n-f_m)) \to_{n,m\to\infty} 0;
\qquad
F = \lim_{n\to\infty} \eta f_n.
\l{2.38}
\eeq
Построение последовательности и доказательство свойств \r{2.38} разобьем
на несколько лемм.

{\bf Лемма 2.14.} {\it Для любой функции $\chi(P,Q)$  из  пространства
Шварца выражение
\bez
\phi(\xi) = \int dPdQ \chi(P,Q) e^{i\Omega(P,Q)} F(\xi)
\eez
определяет основную функцию из $S({\Bbb R}^n)$.
}

{\bf Доказательство.} Аналогично доказательству  леммы  2.9  получаем,
что рассматриваемая функция равна
\bez
\phi(\xi) = \int d\eta F(\eta)
\int dP e^{i\sum_{j=1}^n P_j \frac{\xi_j+\eta_j}{2}} \chi(P,\xi-\eta)
\eez
и принадлежит  $S({\Bbb  R}^n)$  по  известным  свойствам   обобщенных
функций.

{\bf Лемма 2.15} {\it Пусть для функции $F$ выполнено свойство \r{2.36},
а функция $\Phi(\beta,\gamma)$ принадлежит пространству Шварца.  Тогда
интеграл
\bez
f = \int d\beta d\gamma
\Phi(\beta,\gamma)
e^{i \sum_{b=1}^k \beta_b \Omega({\Cal Y}^{(b)})
+ i \sum_{\lambda} \gamma_{\lambda} \Omega({\Cal Z}^{(\lambda)})
} F
\eez
принадлежит классу $S({\Bbb R}^n)$.
}

{\bf Доказательство.} Заметим, что из свойства \r{2.36} вытекает, что
\beq
F =
e^{i
\sum_{a=1}^k \alpha_a \Omega({\Cal X}^{(a)})
} F.
\l{2.39}
\eeq
Пусть $\chi(\alpha)$ --- любая функция $\alpha$,  интеграл от  которой
равен единице. Тогда
\bez
F = \int d\alpha \chi(\alpha)
e^{i
\sum_{a=1}^k \alpha_a \Omega({\Cal X}^{(a)})
} F.
\eez
Отсюда
\bey
f =
\int d\alpha d\beta d\gamma
\chi(\alpha) \Phi(\beta,\gamma)
e^{i \sum_{b=1}^k \beta_b \Omega({\Cal Y}^{(b)})
+ \sum_{\lambda} \gamma_{\lambda} \Omega({\Cal Z}^{(\lambda)})
}
e^{i
\sum_{a=1}^k \alpha_a \Omega({\Cal X}^{(a)})
} F \\
=
\int d\alpha d\beta d\gamma
\chi(\alpha) \Phi(\beta,\gamma)
\\ \times
e^{- \frac{i}{2} \sum_{a=1}^k \alpha_a \beta_a }
e^{
i \sum_{a=1}^k \alpha_a \Omega({\Cal X}^{(a)}) +
i\sum_{b=1}^k \beta_b \Omega({\Cal Y}^{(b)})
+ i\sum_{\lambda} \gamma_{\lambda} \Omega({\Cal Z}^{(\lambda)})
} F.
\eey
По лемме 2.14 получаем, что $f \in S({\Bbb R}^n)$. Лемма доказана.

Подберем последовательность $f_n \in S({\Bbb R}^n,{\Cal L}_k)$ как
\beq
f_n =
\int d\beta d\gamma
\Phi_n(\beta,\gamma)
e^{i \sum_{b=1}^k \beta_b \Omega({\Cal Y}^{(b)})
+ i\sum_{\lambda} \gamma_{\lambda} \Omega({\Cal Z}^{(\lambda)})
} F,
\l{2.40}
\eeq
где
\bez
\Phi_n(\beta,\gamma) = n \Phi(\beta,n\gamma),
\eez
а $\Phi(\beta,\sigma)$  ---  гладкая вещественная функция с компактным
носителем, удовлетворяющая условиям нормировки
\beq
J(2\pi)^k \int d\sigma \Phi(0,\sigma) = 1.
\l{2.41}
\eeq

{\bf Лемма 2.16.} {\it
Последовательность $\eta  f_n$  сходится  к  $F$  в  смысле сходимости
обобщенных функций.
}

{\bf Доказательство.}
Пусть $\varphi \in S({\Bbb R}^n)$. Установим, что
\bez
(\varphi,\eta f_n) \to_{n\to\infty} (\varphi,F),
\eez
где
\bez
\eta f_n = J \int d\alpha
e^{i
\sum_{a=1}^k \alpha_a \Omega({\Cal X}^{(a)})
} f_n.
\eez
Имеем:
\beb
\eta f_n =
J \int d\alpha d\beta d\gamma
\Phi_n(\beta,\gamma)
e^{i
\sum_{a=1}^k \alpha_a \Omega({\Cal X}^{(a)})
}
e^{i \sum_{b=1}^k \beta_b \Omega({\Cal Y}^{(b)})
+ i\sum_{\lambda} \gamma_{\lambda} \Omega({\Cal Z}^{(\lambda)})
} F = \\
J \int d\alpha d\beta d\gamma
\Phi_n(\beta,\gamma)
e^{i
\sum_{a=1}^k \alpha_a \beta_a
}
\\ \times
e^{i \sum_{b=1}^k \beta_b \Omega({\Cal Y}^{(b)})
+ i\sum_{\lambda} \gamma_{\lambda} \Omega({\Cal Z}^{(\lambda)})
}
e^{i
\sum_{a=1}^k \alpha_a \Omega({\Cal X}^{(a)})
}
F = \\
J \int d\alpha d\beta d\gamma
\Phi_n(\beta,\gamma)
e^{i
\sum_{a=1}^k \alpha_a \beta_a
}
e^{i \sum_{b=1}^k \beta_b \Omega({\Cal Y}^{(b)})
+ i\sum_{\lambda} \gamma_{\lambda} \Omega({\Cal Z}^{(\lambda)})
}
F = \\
J (2\pi)^k \int d\gamma \Phi_n(0,\gamma)
e^{i\sum_{\lambda} \gamma_{\lambda} \Omega({\Cal Z}^{(\lambda)})
}
F.
\l{2.42}
\eeb
здесь использованы лемма 2.1 и свойство \r{2.39}. Следовательно,
\beq
(\varphi,\eta f_n) - (\varphi,F) = (\varphi_n,F),
\l{2.43}
\eeq
где
\bez
\varphi_n = \int d\gamma \Phi_n(0,\gamma)
[
e^{-i\sum_{\lambda} \gamma_{\lambda} \Omega({\Cal Z}^{(\lambda)})
}
- 1] \varphi.
\eez
Установим, что последовательность функций $\varphi_n$ сходится к  нулю
в смысле   сходимости   обобщенных   функций.   Для  этого  достаточно
проверить, что   для   любого   конечного   набора   векторов   ${\Cal
W}^{(1)},...,{\Cal W}^{(p)}$ из $\Cal M$ справедливо свойство
\beq
|| \Omega({\Cal W}^{(1)}) ... \Omega({\Cal W}^{(p)}) \varphi_n||_{L^2}
\to_{n\to\infty} 0.
\l{2.44}
\eeq
Заменой $n\gamma = \sigma$ приводим интеграл \r{2.43} к виду
\bey
\varphi_n = \int d\sigma \Phi(0,\sigma)
[
e^{-i \frac{1}{n} \sum_{\lambda} \gamma_{\lambda}
\Omega({\Cal Z}^{(\lambda)})}
- 1] \varphi
\\
= \frac{1}{n} \int d\sigma
\int_0^1 d\tau \Phi(0,\sigma)
(-i\sigma_{\lambda}
\Omega({\Cal Z}^{(\lambda)}))
e^{-i \frac{1}{n} \sum_{\lambda} \gamma_{\lambda}
\Omega({\Cal Z}^{(\lambda)})} \varphi.
\eey
Используя перестановочные  соотношения  из  леммы   2.1,   унитарность
оператора $e^{-i \frac{1}{n} \sum_{\lambda} \gamma_{\lambda}
\Omega({\Cal Z}^{(\lambda)})}$ и компактность носителя функции $\Phi$,
приходим к   свойству  \r{2.44}.  Из  непрерывности функционала
$F$  получаем,  что
последовательность \r{2.43} сходится к нулю. Лемма доказана.

{\bf Лемма 2.17.} {\it При условиях теоремы 2.1 справедливо соотношение
\beq
\rho(\alpha,\beta,\gamma) = \prod_{b=1}^k \delta(\beta_b) R(\gamma),
\l{2.45}
\eeq
где $R(\gamma)$ --- непрерывная в окрестности точки $\gamma=0$ функция.
}

{\bf Доказательство. }  Из свойства \r{2.36} вытекает, что
\bez
e^{i
\sum_{a=1}^k \alpha_a' \Omega({\Cal X}^{(a)})
}
F = F,
\eez
так что
\bey
\rho(\alpha,\beta,\gamma) =
(F,
e^{i
\sum_{a=1}^k \alpha_a' \Omega({\Cal X}^{(a)})
} \\ \times
e^{i
\sum_{a=1}^k \alpha_a \Omega({\Cal X}^{(a)})
+ i\sum_{b=1}^k \beta_b \Omega({\Cal Y}^{(b)})
+ i\sum_{\lambda} \gamma_{\lambda} \Omega({\Cal Z}^{(\lambda)})
}
e^{i
\sum_{a=1}^k \alpha_a'{}' \Omega({\Cal X}^{(a)})
}
F)
\eey
По лемме 2.1 получим
\bey
\rho(\alpha,\beta,\gamma) =
(F,
e^{\frac{i}{2}
\sum_{a=1}^k (\alpha'- \alpha'{}')_a \beta_a } \\
\times
e^{i
\sum_{a=1}^k (\alpha + \alpha' + \alpha'{}')_a \Omega({\Cal X}^{(a)})
+ \sum_{b=1}^k \beta_b \Omega({\Cal Y}^{(b)})
+ \sum_{\lambda} \gamma_{\lambda} \Omega({\Cal Z}^{(\lambda)})
}
F)
\eey
Отсюда получаем,  что обобщенная функция $\rho$ не зависит от $\alpha$
и пропорциональна $\prod_{a=1}^k \delta(\beta_a)$,  то есть
имеет вид \r{2.39}.  Ввиду  непрерывности  функции  \r{2.39}  по  $\gamma$
в окрестности нуля
функция $R(\gamma)$ должна удовлетворять аналогичным свойствам.  Лемма
доказана.

{\bf Лемма 2.18.} {\it
Справделиво свойство
\bez
((f_n,f_m)) \to_{n,m\to\infty} A,
\eez
где $A=const$.
}

{\bf Доказательство.} Ввиду \r{2.42} имеем
\bey
((f_n,f_m)) = (f_n,\eta f_m) = \\
(\int d\beta' d\gamma'
\Phi_n(\beta',\gamma')
e^{- i \sum_{b=1}^k \beta_b' \Omega({\Cal Y}^{(b)})
- i\sum_{\lambda} \gamma_{\lambda}' \Omega({\Cal Z}^{(\lambda)})
} F, \\
J (2\pi)^k \int  d\gamma
\Phi_m(0,\gamma)
e^{i\sum_{\lambda} \gamma_{\lambda} \Omega({\Cal Z}^{(\lambda)})
} F) = \\
J(2\pi)^k \int d\beta' d\gamma' d\gamma \Phi_n(\beta',\gamma')
\Phi_m(0,\gamma) \rho(0,-\beta,\gamma-\gamma') =
\\
J (2\pi)^k \int d\gamma d\gamma' \Phi_n(0,\gamma')
\Phi_m(0,\gamma) R(\gamma - \gamma').
\eey
После замены  $n\gamma'  =  \sigma'$  и  $m\gamma  =  \sigma$   данное
выражение приводится к виду
\bez
((f_n,f_m)) = J(2\pi)^k
\int d\sigma' d\sigma \Phi^*(0,\sigma') \Phi(0,\sigma)
R(\frac{\sigma}{m} - \frac{\sigma'}{n}).
\eez
По теореме Лебега при $n,m\to\infty$
\bez
((f_n,f_m)) \to_{n,m\to\infty}
R(0) J (2\pi)^k |\int d\sigma \Phi(0,\sigma)|^2 = A = const.
\eez
Лемма 2.18 доказана.

Из леммы   2.18   получаем,   что   $\{f_n\}$   ---    фундаментальная
последовательность; действительно,
\bez
((f_n-f_m, f_n-f_m))  =  ((f_n,f_n))  +  ((f_m,f_m))  -  ((f_n,f_m)) -
((f_m,f_n)) \to 0.
\eez
Теорема 2.1 полностью доказана.

{\it Доказательство  леммы  2.10.}   Запишем   равенство   \r{2.18}   в
координатах \r{2.32}.   Пусть  $\alpha_1,...,\alpha_k$  ---  координаты
$X\in {\Cal L}_k$,  $(\beta_1,...,\beta_k)$ --- координаты $Y\in {\Cal
L}_k$. Тогда  инвариантные  относительно  сдвигов  меры интегрирования
$d\mu(X)$ и $d\sigma(Y)$ запишутся как
\bez
d\mu(X) = Jd\alpha_1...d\alpha_k, \qquad
d\sigma(Y) = Kd\beta_1...d\beta_k.
\eez
При этом
\bez
<X,Y> = \sum_{a=1}^k \alpha_a \beta_a.
\eez
Следовательно, левая часть соотношения \r{2.18} запишется как
\bez
JK \int  d\alpha d\beta \rho(\beta) e^{i\sum_{a=1}^k \alpha_a \beta_a}
= JK (2\pi)^k \rho(0),
\eez
что доказывает формулу \r{2.18} при
\beq
\Delta = JK(2\pi)^k.
\l{2.46}
\eeq
Проверим теперь равенство \r{2.19}, которое можно записать в виде
\beq
((F,F))^{\vee} = K\int d\beta \rho(\beta)
(F, e^{i\sum_a \beta_a \Omega({\Cal Y}^{(b)}} F).
\l{2.47}
\eeq
По формуле \r{2.35}
\bez
(F, e^{i\sum_a \beta_a \Omega({\Cal Y}^{(b)}} F) =
J(2\pi)^k \prod_b \delta(\beta_b) ((\overline{f},\overline{f})),
\eez
если $F = \overline{\eta} \overline{f}$.  Следовательно,  правая часть
равенства \r{2.47} имеет вид
\bez
JK (2\pi)^k \rho(0) ((\overline{f},\overline{f}))
\eez
и действительно совпадает с левой частью. Лемма 2.10 доказана.

\section{Гауссовские и квазигауссовские функции}

Гауссовские и квазигауссовские функции часто используются в  квантовой
теории систем со связями. Исследуем их свойства.

\subsection{Формулировка результатов}

Будем называть функцию $f \in S({\Bbb R}^n)$ вида
\beq
f(\xi) = c \exp[\frac{i}{2} \sum_{ij=1}^n \xi_i A_{ij} \xi_j],
\l{3.1}
\eeq
где $c$ --- некоторая константа,  $c\ne 0$,
$A_{ij}$ --- симметричная матрица  с
положительно определенной мнимой частью ($Im A > 0$), {\it гауссовской
функцией.}

Обозначим через  ${\Cal  M}^{\Bbb  C}$  комплексифицированное  фазовое
пространство ---  комплексное  линейное пространство всех наборов $2n$
комплексных чисел    $(P_1,...,P_n;Q_1,...,Q_n)$    с    кососкалярным
произведением \r{2.1}.  В  теории  систем  без связей любой гауссовской
функции можно сопоставить  комплексный  росток  Маслова [10,11] ---  линейное
подпространство $r(A)  \in  {\Cal  M}^{\Bbb  C}$,  состоящее  из  всех
векторов $Y$, удовлетворяющих свойству
\beq
\Omega(Y) f = 0
\l{3.2}
\eeq
(здесь оператор $\Omega(Y)$ имеет вид \r{2.2}).

Для систем  со  связями  равенство  \r{3.2}  можно понимать двояко:  в
смысле пространства $S({\Bbb R}^n)$ и в смысле пространства $\Cal  H$.
Введем поэтому понятия {\it $S$-ростка } и {\it $H$-ростка}.

Назовем
{\it комплексным $S$-ростком Маслова } $r(A)$ множество  всех  $Y  \in
{\Cal M}^{\Bbb  C}$,  таких,  что равенство \r{3.2} выполнено в смысле
пространства $S({\Bbb R}^n)$.

Рассмотрим теперь равенство \r{3.2} в смысле  пространства  $\Cal  H$.
Следует отметить,   что   согласно   лемме  2.5  оператор  $\Omega(Y)$
определен только при
$Y \in ({\Cal L}_k^{\Bbb C})^{< \perp >}$
(здесь
${\Cal L}_k^{\Bbb C}$ --- комплексификация подпространства
${\Cal L}_k \subset {\Cal M}$ фазового пространства).  Кроме  того,  в
пространстве $\Cal  H$  ненулевая  функция  $\Omega(Y)  f$  может быть
эквивалентна нулю,  если имеет нулевую норму.  Приходим  к  следующему
определению.

Назовем {\it комплексным $H$-ростком Маслова} $\check{r}(A)$ множество
всех
$Y \in ({\Cal L}_k^{\Bbb C})^{< \perp >}$,
таких, что
\beq
\Omega(Y) f \sim 0.
\l{3.3}
\eeq

Исследуем свойства $S$-ростка и $H$-ростка.

Многие свойства $S$-ростка аналогичны случаю систем без связей.

{\bf Лемма 3.1.} {\it
1. Комплексный $S$-росток Маслова $r(A)$ является $n$-мерным
подпространством ${\Cal M}^{\Bbb C}$, состоящим из векторов
\bez
r(A) = \{ (P=AQ,Q) | Q \in {\Bbb C}^n \}.
\eez
2. $r(A)$ удовлетворяет свойствам:
\beb
Y \in r(A), \quad Y \ne 0 \Rightarrow \frac{1}{i} <Y,Y^*> > 0.\\
Y_1, Y_2 \in r(A) \Rightarrow <Y_1,Y_2> = 0.
\l{3.4}
\eeb
}

Доказательства лемм приводятся в пунктах 3.2 и 3.3.

Как показывает лемма 3.1,  на $S$-ростке
$r(A)$ можно ввести положительно определенное
скалярное произведение $\frac{1}{i} <Y_1,Y_2^*>$.

Обозначим через  $B:  r(A) \to {\Bbb C}^n$ и $C:  r(A) \to {\Bbb C}^n$
отображения, проектирующие $S$-росток  на  импульсное  и  координатное
подпространства; они сопоставляют вектору $Y = (P,Q) \in r(A)$ векторы
$BY=P$ и $CY=Q$.

{\bf Лемма 3.2.  } {\it Пусть $n$-мерное  подпространство  $r  \subset
{\Cal M}^{\Bbb C}$ удовлетворяет свойствам \r{3.4}. Тогда:

(а) пространство ${\Cal M}^{\Bbb C}$ распадается в прямую сумму
\beq
{\Cal M}^{\Bbb C} = r + r^*;
\qquad r \cap r^* = \{0\}.
\l{3.5}
\eeq

(б) отображение $C:  r \to {\Bbb C}^n$ взаимно однозначно;

(в) $r$   является   комплексным  $S$-ростком  Маслова  для  некоторой
гауссовской функции  \r{3.1},  которая   определяется   однозначно   с
точностью до константы $c$; при этом $A= BC^{-1}$.
}

Используя понятие  $S$-ростка,  можно  получить формулу  для
квадрата нормы   гауссовской   функции   \r{3.1}.   Пусть  $r(A)$  ---
$S$-росток, отвечающий гауссовской  функции  \r{3.1};  рассмотрим  его
подпространство вида
\bez
r_{\perp} (A) = r(A) \cap ({\Cal L}_k^{\Bbb C})^{< \perp >}.
\eez
Через $r_-(A)$  обозначим  ортогональное  дополнение  к $r_{\perp}(A)$
относительно скалярного произведения $\frac{1}{i} <Y_1,Y_2^*>$.

{\bf Лемма 3.3.} {\it
1. Любой вектор $X \in {\Cal L}_k$ однозначно представим в виде суммы
\bez
X = X_- + X_-^*; \quad X_- \in r_-(A);
\eez
2. Справедливы свойства
\bez
dim \quad r_{\perp} (A) = n-k;
\qquad
dim \quad r_- (A) = k.
\eez
3. Отображение  $k$-мерных  пространств  $P_-:  {\Cal  L}_k^{\Bbb C}  \to
r_-(A)$, определенное  по  формуле  $P_-X=X_-$,  является  линейным  и
взаимно однозначным.
}

Определим понятие  якобиана линейного
отображения линейных пространств следующим
образом. Пусть  ${\Cal  L}$  и  ${\Cal  L}'$  ---   два   линейных
$k$-мерных пространства   с  мерами  $d\mu$  и  $d\mu'$,  инвариантные
относительно сдвигов; $P: {\Cal L} \to {\Cal L}'$ --- линейное взаимно
однозначное отображение    этих   пространств.   Пусть   в   некоторых
координатах $(x_1,...,x_k)$ и  $(x_1',...,x_k')$  меры  интегрирования
имеют вид:
\bez
d\mu = J dx_1 ... dx_k; \qquad d\mu' = J' dx_1'...dx_k',
\eez
а оператору   $P$   отвечает   матрица   $P_{ij}$.  Назовем  якобианом
отображения $P$ величину
\beq
\Delta(P) = |det P| \frac{|J'|}{|J|}.
\l{3.6}
\eeq

{\bf Лемма 3.4.} {\it
Определение якобиана \r{3.6} не зависит от выбора системы координат.
}

Сформулируем теперь  результаты,  касающиеся  скалярного  произведения
$((f,f))$.

{\bf Лемма 3.5. } {\it
1. Для $(f,f)$ справедливо равенство
\bez
(f,f) = (2\pi)^{n/2} |c|^2 \Delta (C).
\eez
2. Для $((f,f))$ выполнены соотношения
\beb
((f,f)) = \int_{{\Cal L}_k} d\mu(X) e^{- \frac{1}{2i} <P_-X,(P_-X)^*>}
(f,f); \\
((f,f)) = (2\pi)^{\frac{k+n}{2}} |c|^2
\frac{\Delta (C)}{\Delta (P_-)}.
\l{3.7}
\eeb
}

Из леммы 3.5, в частности, вытекает, что $f \not\sim 0$.

{\it Замечание.}  В  координатной   записи   формула   для   $((f,f))$
существенно усложняется.   Действительно,  если  ввести  базис
${\Cal X}^{(1)}, ..., {\Cal X}^{(k)}$ на ${\Cal L}_k$ (${\Cal X}^{(a)}
= ({\Cal   P}^{(a)},{\Cal   Q}^{(a)})$),   ввести  по  формуле  \r{2.6}
координаты на  ${\Cal  L}_k$,  записать  по   формуле   \r{2.7}   меру
интегрирования и воспользоваться \r{2.8}, то получим
\bez
((f,f)) = (2\pi)^{\frac{k+n}{2}} |c|^2
J \frac{1}{\sqrt{det K}\sqrt{det M}},
\eez
где матрицы $M$ и $K$ имеют вид
\bey
M_{ab} = i \sum_{j=1}^n
[{\Cal P}_j^{(a)} {\Cal Q}_j^{(b)}
- {\Cal Q}_j^{(a)} (A {\Cal Q}^{(b)})_j], \quad a,b = \overline{1,k};\\
K_{js} =
\frac{1}{i} (A-A^*)_{js} +
\sum_{ab=1}^k
[{\Cal P}^{(a)} - A {\Cal Q}^{(a)}]_j
M^{-1}_{ab} [{\Cal P}^{(b)} - A {\Cal Q}^{(b)}]_s.
\eey
Формулы \r{3.7}  значительно  компактнее и имеют явный геометрический
смысл.

Исследуем теперь  свойства  $H$-ростка,  которые   оказываются   менее
тривиальными по сравнению с утверждениями лемм 3.1 и 3.2.

{\bf Теорема 3.1.} {\it 1. Комплексный $H$-росток Маслова
$\check{r}(A)$ является $n$-мерным  подпространством  ${\Cal  M}^{\Bbb
C}$, представимым в виде прямой суммы
\bez
\check{r}(A) = r_{\perp}(A) + {\Cal L}_k^{\Bbb C},
\qquad r_{\perp}(A) \cap {\Cal L}_k^{\Bbb C} = \{ 0\}.
\eez
2. $\check{r}(A)$ удовлетворяет свойствам
\beb
Y \in {\Cal L}_k^{\Bbb C} \quad \Rightarrow \quad Y \in  \check{r}(A);
\quad \frac{1}{i} <Y,Y^*> = 0; \\
Y \in \check{r}(A), \quad Y \notin {\Cal L}_k^{\Bbb C} \quad
\Rightarrow \frac{1}{i} <Y,Y^*> > 0;\\
Y_1,Y_2 \in \check{r}(A) \quad \Rightarrow \quad <Y_1,Y_2> = 0.
\l{3.8}
\eeb
}

Из теоремы 3.1 вытекает важное

{\bf Следствие.} {\it
Пусть $Y \in {\Cal M}$.  Тогда $\Omega(Y)g  \sim  0$  при  всех  $g\in
S({\Bbb R}^n)$ тогда и только тогда, когда $Y\in {\Cal L}_k$.
}

Достаточность доказывалась выше (формула \r{2.11});  для доказательства
необходимости достаточно  заметить,  что $Y \in \check{r}(A)$ при всех
$A$, но,  поскольку $Y=Y^*$,  справедливо свойство $<Y,Y^*> = 0$ и  $Y
\in {\Cal L}_k^{\Bbb C}$.

{\bf Теорема  3.2.}  {\it 1.  Пусть $n$-мерное пространство $\check{r}
\in {\Cal M}^{\Bbb C}$ удовлетворяет свойствам \r{3.8}. Тогда:

(а) справедливы свойства:
\bez
({\Cal L}_k^{\Bbb C})^{<\perp>} = \check{r} + \check{r}^*,
\qquad
\check{r} \cap \check{r}^* = {\Cal L}_k^{\Bbb C}.
\eez
(б) для некоторой гауссовской функции $\check{r} = \check{r}(A)$.

2. Для двух ненулевых гауссовских функций
\bez
f_1(\xi) = c_1 e^{\sum_{ij} \xi_i A^{(1)}_{ij} \xi_j}, \qquad
f_2(\xi) = c_2 e^{\sum_{ij} \xi_i A^{(2)}_{ij} \xi_j}
\eez
комплексные $H$-ростки   Маслова   совпадают   ($\check{r}(A^{(1)})  =
\check{r}(A^{(2)})$) тогда  и  только  тогда,  когда  для   некоторого
числового множителя $c$ справделиво свойство $f_1 \sim c f_2$.
}

Отметим, что  доказательство  второго  утвержения опирается на полноту
квазигауссовских функций (теорема 3.3).

Исследуем теперь обобщенную  функцию  $F=\eta  f  \in  \check{S}({\Bbb
R}^n,{\Cal L}_k)$,  отвечающую  по  формуле \r{2.13} гауссовской функции
\r{3.1}. Оказывается, что если подпространство ${\Cal L}_k$ однозначно
проектируется на координатное подпространство, функция $F$ также имеет
гауссовский вид
\beq
F(\xi) =  \check{c}  e^{\frac{i}{2}  \sum_{ij}  \xi_i   \check{A}_{ij}
\xi_j}.
\l{3.9}
\eeq
При этом число $\check{c}$ и матрицу $\check{A}$ можно выразить  через
$S-$ и $H-$ ростки.

Обозначим через $\check{B}: \check{r}(A) \to {\Bbb C}^n$ и
$\check{C}: \check{r}(A) \to {\Bbb  C}^n$  отображения,  проектирующие
$H$-росток на  импульсное  и  координатное подпространства,  а
через $\Pi:  \check{r}(A) \to  {\Cal  L}_k^{\Bbb  C}$  ---  однозначно
определенный по теореме 3.1 оператор проектирования на
${\Cal  L}_k^{\Bbb  C}$  вдоль  $r_{\perp}(A)$:  он
сопоставляет вектору  $Y  \in   \check{r}(A)$   второе   слагаемое   в
разложении $Y = Y_{\perp} + \Pi Y$, $Y_{\perp} \in r_{\perp}(A)$, $\Pi
Y \in {\Cal L}_k^{\Bbb C}$.

Обозначим через ${\Cal P}_-: \check{r}(A) \to r(A)$ оператор вида:
\beb
{\Cal P}_-Y = Y, \qquad Y \in r_{\perp}(A); \\
{\Cal P}_-Y = P_-Y, \qquad Y \in {\Cal L}_k.
\l{3.10}
\eeb

{\bf Лемма 3.6.} {\it
Пусть подпространство   ${\Cal   L}_k$   однозначно  проектируется  на
координатную плоскость, то есть
\bez
({\Cal P},{\Cal Q}) \in {\Cal L}_k, \quad {\Cal Q}=0 \quad \Rightarrow
\quad {\Cal P} = 0.
\eez
Тогда:

(а) отображение $\check{C}$ взаимно однозначно;

(б)
функция $F(\xi)$ имеет вид \r{3.9}, причем
$\check{A} = \check{B} \check{C}^{-1}$;

(в) для $\check{c}$ справедлива любая из формул:
\beb
\check{c} = c \int_{{\Cal L}_k} d\mu(X)
e^{- \frac{i}{2}<X, C^{-1}\check{C} X>};\\
\check{c} = c (2\pi)^{k/2}
\frac{\sqrt{det[C {\Cal P}_- \check{C}^{-1}]}}{\Delta (P_-)}.
\l{3.11}
\eeb
}

В формулу  \r{3.11} входит квадратный корень из детерминанта оператора
$\Pi \check{C}^{-1} CP_-:  {\Cal L}_k^{\Bbb C} \to  {\Cal  L}_k^{\Bbb  C}$.
Проблема выбора  знака корня (аналогичная проблеме определения индекса
Маслова) может  быть  решена  следующим  образом.  Рассмотрим  частный
случай, когда подпространство ${\Cal L}_k$ состоит из векторов вида $X
= ({\Cal P}=0, {\Cal Q} \in {\Cal L}_k \subset {\Bbb R}^n)$, а матрица
$A$ чисто  мнимая:  $A^* = - A$.  Тогда разложение $X=P_-X + (P_-X)^*$
имеет вид
$$
(0,{\Cal Q}) =
\frac{1}{2} (A{\Cal Q},{\Cal Q}) + \frac{1}{2} (-A{\Cal Q},{\Cal Q}).
$$
Следовательно, $P_-(0,{\Cal  Q})  = \frac{1}{2} (A{\Cal Q},{\Cal Q})$,
так что $CP_-(0,{\Cal Q}) = \frac{1}{2} {\Cal Q}$. Отсюда
$\check{C}^{-1} CP_-  (0,{\Cal Q}) = (0,\frac{1}{2}{\Cal Q})$,  и этот
вектор уже принадлежит ${\Cal L}_k^{\Bbb C}$. Таким образом,
$\Pi \check{C}^{-1}  CP_-  =  \frac{1}{2}$;  проблем  с  выбором знака
квадратного корня не возникает.  В общем же случае ветвь  корня  можно
выбрать по непрерывности.

Проиллюстрируем полученные формулы на одномерном примере.

Пусть $n=1,k=1$,  подпространство  ${\Cal  L}_1$  натянуто  на  вектор
${\Cal X} = ({\Cal  P},{\Cal  Q})$,  ${\Cal  L}_1  =  \{\alpha  ({\Cal
P},{\Cal Q})  \}$,  мера  интегрирования  на  ${\Cal  L}_1$  имеет вид
$d\alpha$. Рассмотрим гауссовскую функцию
\bez
f = c e^{\frac{i}{2} \xi A \xi}.
\eez
Тогда $S$-росток и $H$-росток имеют вид
\bez
r(A) = \{ (A\beta,\beta)\}, \quad
\check{r}(A) = \{ ({\Cal P}\alpha,{\Cal Q}\alpha)\} =
\{\alpha {\Cal X} \},
\eez
так что $\check{A} = {\Cal P}/{\Cal Q}$.
При этом  ортонормированный  базис  на  $S$-ростке  состоит  из одного
вектора
\bez
{\Cal Z} =
\left(\sqrt{\frac{i}{A-A^*}} A, \sqrt{\frac{i}{A-A^*}} \right).
\eez
Поскольку
\bez
{\Cal X} = \lambda {\Cal Z} + \lambda^*{\Cal Z}^*,
\eez
где
\bez
\lambda = \frac{1}{i}
\sqrt{\frac{i}{A-A^*}} ({\Cal P} - A^*{\Cal Q}),
\eez
оператор $P_-$ имеет вид $P_-(\alpha {\Cal X}) = \alpha \lambda  {\Cal
Z}$. Следовательно,
\bez
\Delta (C) = \sqrt{\frac{i}{A-A^*}}, \qquad
\Delta (P_-) = |\lambda|.
\eez
Непосредственное вычисление дает
\bez
\Pi \check{C}^{-1} CP_- {\Cal X} =
\frac{1}{A-A^*} (\frac{\Cal P}{\Cal Q} - A^*)
{\Cal X},
\eez
и по формуле \r{3.11}
\bez
\check{c} =
\frac{c (2\pi)^{k/2}}{|\lambda|}
\sqrt{\frac{1}{A-A^*} (\frac{\Cal P}{\Cal Q} - A^*)}
\eez
Исследуем важный частный случай ${\Cal Q} \to 0$,  ${\Cal P} = const$,
$A=i$. Тогда $|\lambda| \simeq {\Cal P}/\sqrt{2}$,
\bez
\check{c} =
\frac{c(2\pi)^{1/2}}{\sqrt{|{\Cal P}|}}
e^{-i \frac{\pi}{4} sign \frac{\Cal P}{\Cal Q}}
\frac{1}{\sqrt{|{\Cal Q}|}},
\eez
так что
\bez
F \simeq
\frac{c(2\pi)^{1/2}}{\sqrt{|{\Cal P}|}}
e^{-i \frac{\pi}{4} sign \frac{\Cal P}{\Cal Q}}
\frac{1}{\sqrt{|{\Cal Q}|}}
e^{\frac{i}{2} \frac{\Cal P}{\Cal Q} \xi^2}.
\eez
При ${\Cal P}/{\Cal Q} \to 0$ функция $F$ стремится к дельта-функции:
\bez
F \simeq \frac{c}{|{\Cal P}|} \delta(\xi),
\eez
которая оказывается,  таким образом,  предельным  случаем  гауссовской
функции \r{3.b4}.   Из   рассмотренного   примера   видно,   что   для
вырожденного случая,  когда подпространство ${\Cal L}_k$  неоднозначно
проектируется на  координатное  подпространство,  функция $F$ не имеет
вида \r{3.b4}, но может быть представлена как предел функций \r{3.b4}.

Используя гауссовские  функции  $f$  и  опреаторы  $\Omega(Y)$,  можно
построить множества,  плотные  в  $\Cal H$.  Пусть $f$ --- гауссовская
функция \r{3.1}.

{\bf Теорема 3.3.} {\it
1. Множество всех конечных линейных комбинаций функций
\beq
e^{i\Omega(Y)} [f], \qquad Y \in {\Cal L}_k^{<\perp>}
\l{3.12}
\eeq
образует плотное в $\Cal H$ множество.

2. Множество всех конечных линейных комбинаций функций
\beq
\Omega(Y_1^*) ... \Omega(Y_p^*)
[f], \qquad Y_1^*,...,Y_p^* \in \check{r}^*(A).
\l{3.13}
\eeq
плотно в $\Cal H$.
}

Будем называть  конечные  линейные  комбинации  элементов   $\Cal   H$
\r{3.13}, где $p\le s$, {\it квазигауссовскими векторами ранга $s$.} В
дальнейшем нам потребуется следующее утверждение.

{\bf Лемма 3.7}.  {\it На множестве квазигауссовских векторов  $s$-го
ранга справедливы свойства:
\bey
||\Omega(Z) g|| \le \sqrt{s} \sqrt{\frac{1}{i}<Z,Z^*>} ||g||,\\
||\Omega(Z^*) g|| \le \sqrt{s+1} \sqrt{\frac{1}{i}<Z,Z^*>} ||g||,\\
Z \in \check{r}(A).
\eey
}

\subsection{Доказательство свойств комплексного ростка Маслова}

{\it Доказательство леммы 3.1}.  Пусть $Y =  (P,Q)$.  Непосредственным
вычислением получаем, что
\bez
\Omega(Y)f = \sum_{j=1}^n (P-AQ)_j\xi_j f.
\eez
При этом  свойство  \r{3.2}  выполнено  тогда  и  только тогда,  когда
$P=AQ$. Проверим свойства \r{3.4}:
\bez
\frac{1}{i} <Y,Y^*> = \frac{1}{i} \left<
\left( \matrix AQ \\ Q \endmatrix \right),
\left( \matrix A^*Q^* \\ Q^* \endmatrix \right)
\right> =
\frac{1}{i} \sum_{js} Q_j^* (A-A^*)_{js} Q_s > 0.
\eez
\bez
\frac{1}{i} <Y,Y'> = \frac{1}{i} \left<
\left( \matrix AQ \\ Q \endmatrix \right),
\left( \matrix A'Q' \\ Q' \endmatrix \right)
\right> =
\frac{1}{i} \sum_{js} Q_j (A_{sj} - A_{js}) Q_s = 0.
\eez

{\it Доказательство леммы 3.2.} При $Y \in  r  \cap  r^*$  имеем
$<Y,Y^*> =   0$,  что  возможно  лишь  при  $Y=\{0\}$.  Следовательно,
подпространство $r + r^* \subset  {\Cal  M}^{\Bbb  C}$  является
$2n$-мерным, так что $r + r^* = {\Cal M}^{\Bbb C}$.

Предположим, что отображение $C$ не является взаимно однозначным, и $Y
= (P,Q=0) \in r$ для некоторого $P\ne 0$.  Но тогда $<Y,Y^*> =  0$,
что возможно лишь при $P=0$. Противоречие.

Из взаимной однозначности $C$ вытекает, что
\bez
r = \{ (BY,CY) | Y \in r\} =
\{ (BC^{-1}Q,Q) | Q \in {\Bbb C}^n \}.
\eez
С другой стороны, по доказанному в лемме 3.1 $r(A) = \{ (AQ,Q) | Q \in
{\Bbb C}^n\}$.  Отсюда  следует,  что  $r=r(A)$  тогда и только тогда,
когда $A=BC^{-1}$.

{\it Доказательство леммы 3.3.} По лемме  3.2,  любой  вектор  $X  \in
{\Cal L}_k$ однозначно представляется в виде
\beq
X = X_- + X_+, \quad X_- \in r(A), \quad X_+ \in r^*(A).
\l{3.14}
\eeq
Сопрягая это равенство, получим
\bez
X = X_+^* + X_-^*, \quad X_+^* \in r(A), \quad X_-^* \in r^*(A).
\eez
Из однозначности представления \r{3.14} находим,  что $X_+^*  =  X_-$.
Проверим, что $X_- \in r_-(A)$. Для этого достаточно установить, что
\beq
Y \in r_{\perp}(A) \quad \Rightarrow \quad <X_-^*,Y> = 0.
\l{3.15}
\eeq

Для проверки свойства \r{3.15} достаточно заметить,  что из \r{3.14} и
\r{3.4} вытекает $<X,Y> = <X_-^*,Y>$;  свойство же $<X,Y> = 0$ следует
из $Y \in r_{\perp}(A)$. Последнее утверждение леммы проверено.

Определим отображение $P_-$ по формуле $P_-(X+iX') =  X_-+iX_-'$.  Его
линейность очевидна.    Установим    его    взаимную    однозначность.
Предположим, что для некоторых $X,X' \in {\Cal L}_k$ оказалось $X_-  +
iX_-' = 0$. Тогда $X_-^* - iX_-^* = 0$, и
\bez
<X,X'> =  <X_-,X_-^{\prime *}> + <X_-^*,X_-'> = - 2i <X_-',X_-^{\prime
*}>.
\eez
Свойство изотропности $<X,X'> = 0$ может быть выполнено,  если  только
$<X_-',X_-^{\prime *}>  =  0$,  или $X_-' = 0$,  что означает $X_-=0$,
$X=0$, $X'=0$. Тем самым при $X \in {\Cal L}_k^{\Bbb C}$ из $P_-X = 0$
вытекает $X=0$.

Докажем второе  утверждение  леммы.  Пусть  ${\Cal  X}^{(1)},...,{\Cal
X}^{(k)}$ --- базис на ${\Cal L}_k$.  Тогда $Y \in r_{\perp}(A)$ тогда
и только тогда, когда
\beq
\omega_a(Y) = <{\Cal X}^{(a)},Y> = 0.
\l{3.16}
\eeq
Линейные формы  $\omega_1,...,\omega_k$,  зависящие  от  $Y\in  r(A)$,
линейно независимы.  Действительно,  предположим противное.  Тогда для
некоторого $X \in {\Cal L}_k^{\Bbb C}$ $<X,Y> = 0$  для  всех  $Y  \in
r(A)$, что  означает  $<(P_-X)^*,Y>  =  0$  и  $<(P_-X)^*,Y^*>  =  0$,
следовательно, вектор $(P_-X)^*$  косоортогонален  всему  пространству
${\Cal M}^{\Bbb C}$ и равен поэтому нулю. Отсюда вытекает $X=0$.

Достроим набор         $\omega_1,...,\omega_k$        до        базиса
$\omega_1,...,\omega_n$ на  пространстве  линейных   функционалов   на
$r(A)$. Рассмотрим   базис   ${\Cal  Y}^{(1)},...,{\Cal  Y}^{(k)}$  на
$r(A)$, удовлетворяющий   свойству   $\omega_i({\Cal    Y}^{(i)})    =
\delta_{ij}$; тогда   $<{\Cal  X}^{(a)},{\Cal  Y}^{(j)}>=\delta_{aj}$.
Следовательно, подпространство   $r_{\perp}(A)$   является    линейной
оболочкой векторов    ${\Cal   Y}^{(k+1)},...,{\Cal   Y}^{(n)}$.   Оно
$n-k$-мерно; его  ортогональное  дополнение  относительно   скалярного
произведения $\frac{1}{i}     <Y_1,Y_2^*>$     является    $k$-мерным.
Следовательно, $P_-$ --- взаимно  однозначное  отображение  $k$-мерных
пространств. Лемма полностью доказана.

{\it Доказательство леммы 3.4.} При преобразовании координат
\bez
\tilde{x} = Ux, \quad \tilde{x}' = U'x'
\eez
величины $J$ и $J'$ и матрица $P$ переходят в
\bez
\tilde{J} = J |det U|^{-1}, \quad \tilde{J}' = J |det U'|^{-1},
\quad
\tilde{P} = U'PU^{-1}.
\eez
Следовательно,
\bez
\Delta(\tilde{P}) = |det \tilde{P}| \frac{|\tilde{J}'|}{|\tilde{J}|} =
\Delta(P),
\eez
что доказывает корректность определения.

{\it Доказательство леммы 3.5.} Сначала явным вычислением гауссовского
интеграла получим
\bez
(f,f) = (2\pi)^{n/2} |c|^2 \frac{1}{\sqrt{det \frac{A-A^*}{i}}}.
\eez
Введем на $S$-ростке $r(A)$ базис ${\Cal Z}^{(1)},...,{\Cal Z}^{(n)}$,
ортонормированный относительно скалярного произведения
$\frac{1}{i} <Z,Z^*>$. Обозначая
${\Cal Z}^{(\gamma)} \equiv (P^{(\gamma)}_k,Q^{(\gamma)}_k)$, $B_{k\gamma} =
P^{(\gamma)}_k$, $C_{k\gamma}   =   Q^{(\gamma)}_k$,  запишем  условие
ортонормированности в виде
\bez
\frac{1}{i} \sum_{k=1}^n
(P_k^{(\gamma)} Q_k^{(\sigma)*} -
Q_k^{(\gamma)} P_k^{(\sigma)*}) = \delta_{\gamma \sigma},
\eez
или в матричной форме
\bez
\frac{1}{i} (C^+B - B^+C) = 1.
\eez
Поскольку $A=BC^{-1}$, имеем
\bez
A-A^+ = BC^{-1} - (BC^{-1})^+ = C^{-1+}  (C^+B  -  B^+C)  C^{-1}  =  i
(CC^+)^{-1}.
\eez
Из соотношения
\bez
\Delta (C) = |det C|
\eez
вытекает, что
\bez
det \frac{A-A^+}{i} = |det C|^{-2} = [\Delta(C)]^{-2}.
\eez
Отсюда получаем первое утверждение леммы.

Для доказательства  второго утверждения воспользуемся формулой \r{2.5}.
Используя лемму 3.3,  запишем оператор $\Omega(X)$, $X\in {\Cal L}_k$,
в виде
\bez
\Omega(X) = \Omega(P_-X) + \Omega((P_-X)^*).
\eez
Используя формулы из леммы 2.1, запишем
\bez
(f,e^{i\Omega(X)}f) =
(f,
e^{i\Omega((P_-X)^*)}
e^{i\Omega(P_-X)} f)
e^{- \frac{1}{2i} <P_-X, (P_-X)^*>}.
\eez
Поскольку $\Omega(P_-X)f=0$, получим
\bez
(f,e^{i\Omega(X)}f) =
(f,f)
e^{- \frac{1}{2i} <P_-X, (P_-X)^*>}.
\eez
Приходим к  первой  формуле  \r{3.7}.  Вычисляя гауссовский интеграл,
получаем вторую формулу.

{\it Доказательство теоремы 3.1} разобьем на несколько лемм.

{\bf Лемма 3.8.} {\it
Справедливы свойства  $r_{\perp}(A)  \subset  \check{r}(A)$  и  ${\Cal
L}_k^{\Bbb C} \subset \check{r}(A)$.
}

{\it Доказательство.} Пусть $Y \in  r_{\perp}(A)$.  Тогда  соотношения
\r{3.2} и \r{3.3} выполнены,  и $Y \in \check{r}(A)$.  Первое свойство
проверено. Второе   свойство   вытекает   из   установленного    ранее
соотношения \r{2.11}.

{\bf Лемма 3.9.} {\it Справедливы свойства:
\beb
Y_1,Y_2 \in \check{r}(A) \quad \Rightarrow \quad <Y_1,Y_2> = 0,\\
Y \in \check{r}(A), Y\in \check{r}^*(A) \quad
\Rightarrow \quad \frac{1}{i} <Y,Y^*> = 0,\\
Y \in \check{r}(A), Y\notin \check{r}^*(A) \quad
\Rightarrow \quad \frac{1}{i} <Y,Y^*> > 0.
\l{3.17}
\eeb
}

{\it Доказательство.} Пусть $Y_1,Y_2 \in \check{r}(A)$. Тогда
\bez
\Omega(Y_1) \Omega(Y_2) f \sim 0, \qquad
\Omega(Y_2) \Omega(Y_1) f \sim 0.
\eez
Отсюда
\bez
[\Omega(Y_1),\Omega(Y_2)] f \sim 0,
\eez
что означает $<Y_1,Y_2> = 0$,  поскольку $f\sim 0$. Первое утверждение
леммы доказано.

При $Y  \in  \check{r}(A)$  и  $Y^*\in  \check{r}(A)$  (так как $Y \in
\check{r}^*(A)$) отсюда получаем $<Y,Y^*> = 0$.

Пусть, наконец,  $Y\in \check{r}(A)$,  но $Y^*  \notin  \check{r}(A)$.
Тогда $\Omega(Y^*) f \not\sim 0$, что означает
\bez
((\Omega(Y^*)f, \Omega(Y^*)f)) > 0,
\eez
или
\bez
((f,\Omega(Y) \Omega(Y^*) f)) > 0.
\eez
Отсюда
\bez
((f,[\Omega(Y) \Omega(Y^*)] f)) > 0,
\eez
что означает $\frac{1}{i} <Y,Y^*> > 0$. Лемма доказана.

{\bf Лемма 3.10.} {\it
1. Пересечение  $r_{\perp}(A)  \cap  {\Cal  L}_k^{\Bbb  C}$ состоит из
одного нуля.

2. Пространство $\check{r}(A)$ не более чем $n$-мерно.

3. Справедливо разложение в прямую сумму
\bez
\check{r}(A) = r_{\perp} + {\Cal L}_k^{\Bbb C}.
\eez
4. Пересечение $\check{r}(A) \cap \check{r}^*(A)$ совпадает  с  ${\Cal
L}_k^{\Bbb C}$.
}

{\it Доказательство.} Пусть $Y \in {\Cal L}_k$.  Тогда  по  лемме  3.8
имеем $Y\in  \check{r}(A)$  и  $Y^* \in \check{r}(A)$.  Следовательно,
$<Y,Y^*> = 0$ по лемме 3.9. С другой стороны, при $Y \in r_{\perp}(A)$
равенство $<Y,Y^*>  =  0$  может выполняться только при $Y=0$.  Первое
утверждение леммы  доказано.  Из  него  вытекает,   что   пространство
$r_{\perp}(A) +  {\Cal  L}_k^{\Bbb  C}  \subset \check{r}(A)$ является
$n$-мерным. Введем на нем базис ${\Cal  Z}^{(1)},...,{\Cal  Z}^{(n)}$.
По лемме  3.9  получим,  что  для  всех $Y \in \check{r}(A)$ выполнены
соотношения
\beq
<Z^{(\gamma)},Y> = 0, \qquad \gamma = \overline{1,n}.
\l{3.18}
\eeq
Но $n$    уравнений   \r{3.18}   задают   $n$-мерное   подпространство
$2n$-мерного пространства   ${\Cal   M}^{\Bbb   C}$.   Поэтому    $dim
\check{r}(A) \le n$. Поскольку $n$-мерное подпространство
$r_{\perp}(A) +  {\Cal  L}_k^{\Bbb  C}   \subset   \check{r}(A)$   уже
построено, оно совпадает с $\check{r}(A)$. Второе и третье утверждение
доказаны.

Докажем четвертое утверждение.  Включение ${\Cal L}_k^{\Bbb C} \subset
\check{r}(A) \cap  \check{r}^*(A)$  вытекает  из  леммы 3.8.  Проверим
обратное. Пусть $Y \in \check{r}(A)$ и $Y\in \check{r}^*(A)$. Тогда по
лемме 3.9
\bez
<Y,Y^*> = 0.
\eez
Запишем $Y  =  X  +  Y_{\perp}$,  где  $X  \in  {\Cal  L}_k^{\Bbb C}$,
$Y_{\perp} \in r_{\perp}(A)$. Тогда
\bez
<Y,Y^*> = <Y_{\perp},Y_{\perp}^*>.
\eez
Следовательно, $<Y_{\perp},Y_{\perp}^*> = 0$,  что означает $Y_{\perp}
= 0$, или $Y \in {\Cal L}_k^{\Bbb C}$. Лемма полностью доказана.

Из лемм 3.8 -- 3.10 вытекает утверждение теоремы 3.1.

{\it Доказательство первого утверждения теоремы 3.2} также разобьем на
несколько лемм.

{\bf Лемма 3.11.} {\it
Пусть $n$-мерное подпространство $\check{r} \subset {\Cal M}^{\Bbb C}$
удовлетворяет свойствам \r{3.8}. Тогда:

1. Справедливо  свойство  $\check{r}  +  \check{r}^*   \subset   {\Cal
L}_k^{\Bbb C}$.

2. Пересечение   $\check{r}   \cap  \check{r}^*$  совпадает  с  ${\Cal
L}_k^{\Bbb C}$.

3. Пространство $\check{r} + \check{r}^*$ не менее чем $2n-k$-мерно.
}

{\it Доказательство.}   Поскольку   ${\Cal   L}_k^{\Bbb   C}   \subset
\check{r}$ по первому свойству \r{3.8},  из третьего свойства  \r{3.8}
получаем, что   $\check{r}  \subset  (\Cal  L_k^{\Bbb  C})^{<\perp>}$.
Аналогично, $\check{r}^*  \subset  (\Cal   L_k^{\Bbb   C})^{<\perp>}$.
Отсюда получаем первое утверждение.

Для доказательства второго утверждения заметим,  что
${\Cal L}_k^{\Bbb C} \subset \check{r}$,
${\Cal L}_k^{\Bbb C} \subset \check{r}^*$, поэтому
${\Cal L}_k^{\Bbb C} \subset  \check{r}  \cap  \check{r}^*$.  Проверим
обратное включение.  При $Y \in \check{r}$ и $Y^* \in \check{r}$ имеем
$<Y,Y^*> = 0$. Следовательно, по второму свойству \r{3.8} $Y \in {\Cal
L}_k^{\Bbb C}$. Второе утверждение доказано.

Для проверки     третьего     утверждения    введем    базис    ${\Cal
X}^{(1)},...,{\Cal X}^{(k)}$ на ${\Cal L}_k$. Дополним его до базисов
$({\Cal X}^{(1)},...,{\Cal    X}^{(k)},    {\Cal    Y}^{(1)},...,{\Cal
Y}^{(n-k)})$ на $\check{r}$ и
$({\Cal X}^{(1)},...,{\Cal    X}^{(k)},    {\Cal    Y}^{(1)*},...,{\Cal
Y}^{(n-k)*})$ на $\check{r}^*$.
Покажем, что векторы
${\Cal X}^{(1)},...,{\Cal    X}^{(k)},    {\Cal     Y}^{(1)},...,{\Cal
Y}^{(n-k)}, {\Cal   Y}^{(1)*},...,{\Cal   Y}^{(n-k)*}$,  принадлежащие
$\check{r} + \check{r}^*$,  линейно независимы. Предположим противное;
тогда для некоторых
\bez
Y_- =  \sum_a  \alpha_a {\Cal X}^{(a)} + \sum_{\gamma} \beta_{\gamma}
{\Cal Y}^{(\gamma)} \in \check{r},
\qquad
Y_+ = \sum_{\gamma} \beta_{\gamma}' {\Cal Y}^{(\gamma)*}
\eez
выполнялось свойство $Y_- + Y_+ = 0$. Отсюда $Y_+ \in {\Cal L}_k^{\Bbb
C}$, что   означает   $Y_+=0$.  Следовательно,  $Y_-=0$.  Из  линейной
независимости системы  $2n-k$  векторов  вытекает  третье  утверждение
леммы.

Поскольку $\check{r}   +   \check{r}^*   \subset   ({\Cal   L}_k^{\Bbb
C})^{<\perp>}$ ---  не   менее   чем   $2n-k$-мерное   подпространство
$2n-k$-мерного пространства, из леммы 3.11 вытекает

{\bf Следствие.} ${\Cal L}_k^{\Bbb C} = \check{r} + \check{r}^*$.

Таким образом, пункт 1а теоремы 3.2 доказан.

{\bf Лемма 3.12. } {\it
Пусть $n$-мерное подпространство $\check{r} \subset {\Cal M}^{\Bbb C}$
удовлетворяет свойствам \r{3.8}.  Тогда $\check{r} = \check{r}(A)$ для
некоторой гауссовской функции.
}

{\it Доказательство.}  Построим  искомую гауссовскую функцию следующим
образом. Пусть ${\Cal G}_k$ ---  калибровочная  плоскость  для  ${\Cal
L}_k$. Тогда  по  лемме 2.8 любой элемент $Z \in \check{r}$ однозначно
представляется в виде суммы
\bez
Z = X + Y + Z_{\perp},
\eez
где $X  \in  {\Cal  L}_k^{\Bbb  C}$,  $Y  \in  {\Cal  G}_k^{\Bbb  C}$,
$Z_{\perp} <\perp> {\Cal L}_k^{\Bbb C}$ и
$Z_{\perp} <\perp> {\Cal G}_k^{\Bbb C}$.
При этом  $<X',Z>=0$  для  всех  $X' \in {\Cal L}_k^{\Bbb C}$;  отсюда
$<X',Y> = 0$ и $Y=0$. Следовательно,
\bez
Z = X + Z_{\perp}.
\eez
Тем самым $H$-росток $\check{r}$ распадается в прямую сумму
\bez
\check{r} = {\Cal L}_k^{\Bbb C} + \check{r}_{\perp},
\eez
причем $\check{r}_{\perp}$ косоортогонально
${\Cal G}_k^{\Bbb C}$, а скалярное произведение
$\frac{1}{i} <Z,Z^*>$ на $\check{r}_{\perp}$ положительно  определено.
Введем на  ${\Cal L}_k$ базис ${\Cal X}^{(1)},...,{\Cal X}^{(k)}$,  на
$\check{r}_{\perp}$ --- ортонормированный базис
${\Cal Z}^{(1)},...,{\Cal Z}^{(n-k)}$,
на ${\Cal G}_k$ --- базис
${\Cal Y}^{(1)},...,{\Cal Y}^{(k)}$,  удовлетворяющий свойству \r{2.16}.
Тогда
\bey
<{\Cal X}^{(a)},{\Cal X}^{(b)}> = 0,
\quad
<{\Cal Y}^{(a)},{\Cal Y}^{(b)}> = 0,
\\
<{\Cal X}^{(a)},{\Cal Y}^{(b)}> = \delta_{ab},
\quad
<{\Cal X}^{(a)},{\Cal Z}^{(\rho)}> = 0,
\\
<{\Cal Y}^{(a)},{\Cal Z}^{(\rho)}> = 0,
\quad
\frac{1}{i}
<{\Cal Z}^{(\rho)},{\Cal Z}^{(\sigma)*}> = \delta_{\rho\sigma}.
\eey
Построим $n$-мерное пространство $r$, натянутое на векторы
${\Cal Z}^{(1)},...,{\Cal Z}^{(n-k)}$ и
${\Cal W}^{(1)},...,{\Cal W}^{(k)}$ вида
${\Cal W}^{(a)} = \frac{1}{\sqrt{2}}
({\Cal X}^{(a)} - i {\Cal Y}^{(a)})$. Тогда
\bez
\frac{1}{i} <{\Cal W}^{(a)}, {\Cal W}^{(b)*}> = \delta_{ab},
\qquad
<{\Cal W}^{(a)}, {\Cal Z}^{(\rho)*}> = 0.
\eez
Следовательно, $r$ удовлетворяет условиям леммы 3.2  и  для  некоторой
гауссовской функции \r{3.1} является $S$-ростком
\bez
r = r(A).
\eez
при этом  $r_{\perp}(A)  =  \check{r}$;  следовательно,  $\check{r}  =
\check{r}(A)$. Лемма 3.12 доказана.

Первое утверждение   теоремы   3.2   доказано.   Второе    утверждение
доказывается в пункте 3.3.

{\it Доказательство леммы 3.6.}

Для проверки    взаимной    однозначности    отображения   $\check{C}$
предположим противное. Именно, пусть вектор
$Y = \left( \matrix P \\ 0 \endmatrix \right)$
принадлежит $\check{r}(A)$. Тогда $<Y,Y^*>=0$, и по теореме 3.1 $Y \in
{\Cal L}_k^{\Bbb C}$, что противоречит условию леммы 3.6.

Для проверки   формулы  \r{3.9}  удобно  записать  формулу  \r{2.13}  в
координатном виде. Пусть ${\Cal X}^{(1)},...,{\Cal X}^{(k)}$ --- базис
на ${\Cal  L}_k$,  $\alpha_1,...,\alpha_k$  ---  координаты  на ${\Cal
L}_k$ вида
\beq
X = \sum_a \alpha_a {\Cal X}^{(a)}, \qquad X \in {\Cal L}_k,
\l{3.19}
\eeq
а
\bez
d\mu(X) = Jd\alpha_1...d\alpha_k.
\eez
Тогда
\beq
F = J  \int  d\alpha_1...d\alpha_k
e^{i\sum_a  \alpha_a  \Omega({\Cal X}^{(a)})} f.
\l{3.20}
\eeq
При ${\Cal  X}^{(a)}  =  ({\Cal  P}^{(a)},{\Cal  Q}^{(a)})$ и $f(\xi)$
гауссовского вида \r{3.1} воспользуемся формулой \r{2.3}:
\bey
(e^{i\sum_a  \alpha_a  \Omega({\Cal X}^{(a)})} f)(\xi)
=
e^{i\sum_{ai} \alpha_a {\Cal P}^{(a)}_i \xi_i
- \frac{i}{2} \sum_{abi} \alpha_a \alpha_b
{\Cal P}^{(a)}_i  {\Cal  Q}^{(b)}_i}
f(\xi  -  \sum_a  \alpha_a {\Cal Q}^{(a)})
= \\
e^{i\sum_{ai} \alpha_a {\Cal P}^{(a)}_i \xi_i
- \frac{i}{2} \sum_{abi} \alpha_a \alpha_b
{\Cal P}^{(a)}_i  {\Cal  Q}^{(b)}_i}
c
e^{\frac{i}{2} \sum_{ij}
(\xi_i  -  \sum_a  \alpha_a {\Cal Q}^{(a)}_i)
(\xi_j  -  \sum_b  \alpha_b {\Cal Q}^{(b)}_j)
}
\eey
Непосредственное вычисление гауссовского интеграла \r{3.20} дает
\beq
F(\xi) = \frac{Jc (2\pi)^{k/2} }{\sqrt{det M}} e^{\frac{i}{2}
\sum_{ij} \xi_i \check{A}_{ij} \xi_j},
\l{3.21}
\eeq
где
\bez
\check{A}_{ij} = A_{ij} + \sum_{ab}
({\Cal P}^{(a)} - A{\Cal Q}^{(a)})_i
M^{-1}_{ab} ({\Cal P}^{(b)} - A{\Cal Q}^{(b)})_j,
\eez
\bez
M_{ab} = i \sum_j
{\Cal P}^{(a)}_j {\Cal Q}^{(b)}_j - i \sum_{ij}
{\Cal Q}^{(a)}_i A_{ij} {\Cal Q}^{(b)}_j.
\eez
Следовательно, функция $F$ действительно имеет вид \r{3.9}.

Указанные вычисления корректны, если только матрица $M_{ab}$ обратима.
Докажем это.  Предположим  противное:  пусть  для  некоторого   набора
$\alpha_1,..,\alpha_k$ справедливо     свойство     $\sum_b     M_{ab}
\alpha_b=0$. Тогда вектор
\bez
\left( \matrix
\sum_b A{\Cal Q}^{(b)} \alpha_b \\
\sum_b {\Cal Q}^{(b)} \alpha_b
\endmatrix \right),
\eez
принадлежащий $r(A)$,    косоортогонален    ${\Cal   L}_k^{\Bbb   C}$,
Следовательно, вектор \r{3.f3} принадлежит $r_{\perp}(A)$, а значит, и
$\check{r}(A)$. Далее, вектор
\bez
\left( \matrix
\sum_b {\Cal P}^{(b)} \alpha_b \\
\sum_b {\Cal Q}^{(b)} \alpha_b
\endmatrix \right)
\eez
также принадлежит ${\Cal L}_k^{\Bbb C}$,  а значит,  и $\check{r}(A)$.
Отсюда
\bez
\left( \matrix
\sum_b ({\Cal P}^{(b)} - A{\Cal Q}^{(b)}) \alpha_b \\
0
\endmatrix \right) \in \check{r}(A)
\eez
Но по доказанному  выше  свойству  обратимости  оператора  $\check{C}$
такое невозможно. Следовательно, матрица $M$ обратима.

Проверим свойство $\check{A} = \check{B}\check{C}^{-1}$. Пусть $Y =
\left( \matrix P\\Q \endmatrix \right) \in  \check{r}(A)$.  Тогда
$\Omega(Y) f \sim 0$, и для любого $g$ справедливо свойство
\bez
((g, e^{i\Omega(Y)} f)) = ((g,f)),
\eez
или ввиду унитарности оператора $e^{i\Omega(Y)}$
\beq
((e^{-i\Omega(Y)}g, f)) = ((g,f)).
\l{3.22}
\eeq
Свойство \r{3.22} означает, что
\beq
\int d\xi [e^{-i\Omega(P,Q)} g]^*(\xi) F(\xi) =
\int d\xi g^*(\xi) F(\xi), \qquad (P,Q) \in \check{r}(A).
\l{3.23}
\eeq
По формуле \r{2.3} свойство \r{3.23} приводится к виду
\bez
\int d\xi e^{i\sum_j [P_j\xi_j + \frac{1}{2} P_jQ_j]}
g^*(\xi+Q) F(\xi) = \int d\xi g^*(\xi) F(\xi).
\eez
Заменой $\xi + Q \Rightarrow \xi$ отсюда получим:
\bez
\int d\xi g^*(\xi)
e^{i\sum_j [P_j\xi_j - \frac{1}{2} P_jQ_j]}
F(\xi - Q) = \int d\xi g^*(\xi) F(\xi),
\eez
или
\beq
e^{i\sum_j [P_j\xi_j - \frac{1}{2} P_jQ_j]}
F(\xi - Q) = F(\xi).
\l{3.24}
\eeq
Для гауссовской  функции  $F$  \r{3.9}  равенство \r{3.24} выполнено,
если только
\beq
\check{A} Q = P, \qquad (P,Q) \in \check{r}(A).
\l{3.25}
\eeq
Так как  $P=\check{B}\alpha$,  $Q=\check{C}\alpha$,  свойство \r{3.25}
означает, что $\check{A} = \check{B}\check{C}^{-1}$.

Докажем третье утверждение леммы.  Предэкспоненту в  формуле  \r{3.21}
можно записать как
\beq
\check{c} = Jc \int d\alpha_1 ... d\alpha_k
e^{- \frac{1}{2} \sum_{ab} \alpha_a M_{ab} \alpha_b}.
\l{3.26}
\eeq
Запишем интеграл  \r{3.26}  в бескоординатном виде.  Вновь перейдем от
переменных $\alpha$ к переменным $X \in {\Cal L}_k$ \r{3.19}. Заметим,
что
\bey
\sum_{ab} \alpha_a M_{ab} \alpha_b =
\sum_{ab}
i\alpha_a [\sum_j {\Cal P}_j^{(a)} {\Cal Q}_j^{(b)} -
\sum_{ij} {\Cal Q}_i^{(a)} A_{ij} {\Cal Q}_j^{(b)}] \alpha_b \\
=
i <X, C^{-1} \check{C} X>.
\eey
Отсюда получаем первую из формул \r{3.11}.

Для получения второй формулы заметим, что, с одной стороны,
\bez
((f,f)) =  (f,F)  =  \int  d\xi  c^*  e^{\frac{i}{2}  \sum_{ij}  \xi_i
(\check{A} - A^*)_{ij} \xi_j} = c^*\check{c}
\frac{(2\pi)^{n/2}}{\sqrt{det (\frac{\check{A} - A^*}{i})}},
\eez
а с  другой  стороны,  справедлива  формула  \r{3.7}.  Сравнивая  эти
соотношения, получаем, что должно быть
\beq
\check{c} = (2\pi)^{k/2} c
\frac{\Delta(C)}{\Delta(P_-)}
\sqrt{det(\frac{\check{A} - A^*}{i})}.
\l{3.27}
\eeq

Упростим выражение     \r{3.27}.     Введем     на      $r_{\perp}(A)$
ортонормированный базис   ${\Cal  Z}^{(k+1)},...,{\Cal  Z}^{(n)}$,  на
$r_-(A)$ ---   ортонормированный   базис   ${\Cal   W}^{(1)},...,{\Cal
W}^{(k)}$, на   ${\Cal   L}_k$  ---  базис  ${\Cal  X}^{(1)},...,{\Cal
X}^{(k)}$. Тогда можно записать
\bez
{\Cal W}^{(a)} =
\left( \matrix
B_{ia} \\ C_{ia}
\endmatrix
\right), \quad
{\Cal Z}^{(\alpha)} =
\left( \matrix
B_{i\alpha} \\ C_{i\alpha}
\endmatrix
\right) =
\left( \matrix
\check{B}_{i\alpha} \\ \check{C}_{i\alpha}
\endmatrix
\right),
\quad
{\Cal X}^{(a)} =
\left( \matrix
\check{B}_{ia} = {\Cal P}^{(a)}_i \\
\check{C}_{ia} = {\Cal Q}^{(a)}_i
\endmatrix
\right),
\eez
где $B_{is}$,   $C_{is}$,   $\check{B}_{is}$,   $\check{C}_{is}$   ---
матрицы, отвечающие   операторам   $B,C:   r(A)   \to   {\Bbb   C}^n$,
$\check{B},\check{C}:\check{r}(A) \to {\Bbb C}^n$.

Поскольку $A=BC^{-1}$, $\check{A} = \check{B}\check{C}^{-1}$, имеем:
\bez
\frac{1}{i} (\check{A} - A^*) =
\frac{1}{i} (\check{B} \check{C}^{-1} - (C^{+})^{-1} B^+) =
(C^+)^{-1} \frac{1}{i} (C^+\check{B} - B^+\check{C}) C^{-1}.
\eez
Обозначая через ${\Cal P}_-$ матрицу
\beq
{\Cal P}_- = \frac{1}{i} (C^+\check{B} - B^+\check{C}),
\l{3.28}
\eeq
учитывая, что $\Delta(C) = |det C|$, из \r{3.27} получаем:
\beq
\check{c} = (2\pi)^{k/2} \frac{c}{\Delta(P_-)}
\sqrt{det (\check{C}^{-1} C {\Cal P}_-)}.
\l{3.29}
\eeq
Исследуем матрицу ${\Cal P}_-$ \r{3.28} размера $n\times n$. Имеем:
\bey
({\Cal P}_-)_{ab} = \frac{1}{i} <{\Cal X}^{(b)}, {\Cal W}^{(a)*}>,
\quad
({\Cal P}_-)_{a\nu} = \frac{1}{i} <{\Cal Z}^{(\nu)}, {\Cal W}^{(a)*}> =
0,\\
({\Cal P}_-)_{\mu b} = \frac{1}{i} <{\Cal X}^{(b)}, {\Cal Z}^{(\mu)*}>
= 0, \quad
({\Cal P}_-)_{\mu\nu} = \frac{1}{i} <{\Cal Z}^{(\nu)}, {\Cal Z}^{(\mu)*}> =
\delta_{\mu\nu},
\eey
здесь $\mu,\nu = \overline{k+1,n}$, $a,b=\overline{1,k}$.

Таким образом, матрица ${\Cal P}_-$ имеет обычный вид
\bez
{\Cal P}_- = \left( \matrixx
P_- & 0 \\ 0 & 1 \endmatrixx \right).
\eez
При этом $P_-$ --- матрица размером $k\times k$,  отвечающая оператору
$P_-: {\Cal L}_k \to r_-(A)$, так как
\bez
P_- (\sum_b \alpha_b {\Cal X}^{(b)})  =  \sum_{ab}  ({\Cal  P}_-)_{ab}
\alpha_b {\Cal W}^{(a)}.
\eez
Что касается  оператора  $\check{C}C:  r(A)  \to \check{r}(A)$,  то он
является тождественным     на     подпространстве      $r_{\perp}(A)$;
следовательно, его матрица имеет блочный вид
\bez
\check{C}^{-1} C =
\left( \matrixx
S & 0 \\
S' & 1
\endmatrixx \right),
\eez
так что
\bez
\check{C}^{-1} C {\Cal P}_- =
\left( \matrixx
SP_- & 0 \\
S'P_- & 1
\endmatrixx \right),
\eez
и $det(\check{C}^{-1}C{\Cal   P}_-)   =   det(SP_-)$.  Матрица  $SP_-$
соответствует оператору $\Pi \check{C}^{-1} CP_-$.
Лемма доказана.

\subsection{Доказательство  свойства полноты системы квазигауссовских функций
и его следствий}

Докажем теорему 3.3 и второе утверждение теоремы 3.2.

{\it Доказательство теоремы 3.3} разобьем на несколько лемм.

{\bf Лемма 3.13.} {\it
Пусть ${\Cal   Z}   \in   {\Cal   M}^{\Bbb  C}$.  Тогда  функция  вида
$e^{i\Omega(Z)}f$ эквивалентна   $ce^{i\Omega(Y)}f$,   где   $c$   ---
числовой множитель, $Y\in {\Cal L}_k^{<\perp>}$.
}

{\it Доказательство.}  Рассмотрим  сначала  случай,   когда   $Z   \in
r_-^*(A)$. В  этом  случае  $Z_-^*  \in  r_-(A)$,  и  по лемме 3.3 для
некоторых $X \in {\Cal L}_k^{\Bbb C}$ и $X_-^* \in r_-(A)$ справедливо
свойство
\bez
X = Z^* + X_-^*.
\eez
Тем самым
\bez
\Omega(Z) = \Omega(X^* - X_-).
\eez
По лемме 2.1
\bez
e^{i\Omega(Z)} f = e^{i\Omega(X^*)} e^{-i\Omega(X_-)}
e^{\frac{i}{2} <X^*,X_->} f.
\eez
Поскольку $e^{-i\Omega(X_-)} f = f$ при $X_-\in r_-(A)$, а
$e^{i\Omega(X^*)} f  \sim  f$  при  $X^* \in {\Cal L}_k^{\Bbb C}$,
имеем:
$e^{i\Omega(Z)} f \sim c_1 f$ при
$c_1 = e^{\frac{i}{2} <X^*,X_->}$.

Пусть теперь $Z \in r^*(A)$. Тогда можно записать
\bez
Z = Z_- + Z_{\perp},  \qquad Z_- \in  r_-^*(A),  \quad  Z_{\perp}  \in
r_{\perp}^*(A).
\eez
Следовательно,
\beq
e^{i\Omega(Z)} f = e^{i\Omega(Z_-)} e^{i\Omega(Z_{\perp})} f
\sim c_1 e^{i\Omega(Z_+)} f.
\l{3.30}
\eeq
Рассмотрим функцию
\bez
e^{i\Omega(Z_{\perp} + Z_{\perp}^*} f =
e^{i\Omega(Z_{\perp})}
e^{i\Omega(Z^*_{\perp})}
e^{-\frac{i}{2} <Z_{\perp},Z_{\perp}^*>} f.
\eez
Поскольку $e^{i\Omega(Z_{\perp}^*)}  f  =  f$,  она  с  точностью   до
константы совпадает  с  \r{3.30}.  Тем  самым  для  некоторой  $c_2$ и
$Y=Z_{\perp} + Z_{\perp}^*$
\beq
e^{i\Omega(Z)} f  \sim  c_2  e^{i\Omega(Y)}  f,  \qquad  Y  \in  {\Cal
L}_k^{<\perp>}.
\l{3.31}
\eeq

Рассмотрим теперь   общий  случай.  Для  произвольного  $Z  \in  {\Cal
M}^{\Bbb C}$ можно записать разложение
\bez
Z = Z_+ + Z_-, \qquad Z_+ \in r^*(A), \quad Z_-\in r(A),
\eez
и
\bey
e^{i\Omega(Z)} f = e^{i\Omega(Z_++Z_-)} f =
e^{i\Omega(Z_+)} e^{i\Omega(Z_-)} e^{-\frac{i}{2} <Z_+,Z_->} f
\\ =
e^{i\Omega(Z_+)} e^{-\frac{i}{2} <Z_+,Z_->} f.
\eey
Функция $e^{i\Omega(Z_+)}f$, как установлено выше, имеет вид \r{3.31}.
Утверждение леммы полностью доказано.

{\bf Лемма 3.14.} {\it
Совокупность линейных комбинаций элементов $\Cal H$
\bez
[e^{i\Omega(P,Q)} f]
\eez
образует всюду плотное множество в $\Cal H$.
}

{\it Доказательство.} Рассмотрим интеграл вида
\beq
\int dP\chi(P) e^{i\Omega(P,0)}f = \int
dP\chi(P) e^{i\sum_j P_j\xi_j} f,
\l{3.32}
\eeq
где $\chi \in S({\Bbb R}^n)$.
Покажем, что  он  может  быть  приближен  с  любой точностью конечными
линейными комбинациями функций
$e^{i\Omega(P_s,0)}f$. Действительно,  рассмотрим  достаточно  большой
шар объема $V$,  интеграл по которому отличается от интеграла по всему
$n$-мерному пространству   \r{3.32}   на  величину,  не  превосходящую
${\varepsilon}/2$ по норме.  Для каждой точки $P$ выберем шарообразную
окрестность $O_P$ объема $V(O_P)$,  удовлетворяющую свойству: при всех
$P' \in O_P$ справедливо соотношение
\bez
|| \chi(P) e^{i\Omega(P,0)}f -
\chi(P') e^{i\Omega(P',0)}f|| < \frac{{\varepsilon}}{2V}.
\eez
Существование такой  окрестности  вытекает  из  сильной  непрерывности
функции $\chi(P) e^{i\Omega(P,0)}f$. Из этого покрытия шара объема $V$
выберем конечное подпокрытие,  состоящее  из  окрестностей  $O_{P_j}$.
Рассмотрим сумму
\beq
\sum_j V(O_{P_j}) \chi(P_j) e^{i\Omega(P_j,0)} f;
\l{3.33}
\eeq
ее отличие от интеграла \r{3.32} по объему $V$ равно
\bez
\sum_j \int_{O_{P_j}} dP
[\chi(P_j) e^{i\Omega(P_j,0)} f -
\chi(P) e^{i\Omega(P,0)} f]
\eez
и не превосходит ${\varepsilon}/2$ по норме.  Тем самым сумма \r{3.33}
может аппроксимировать  интеграл  \r{3.32} с любой точностью по норме.
Следовательно, функцию  $\tilde{\chi}(\xi)  f(\xi)$,  где  $\chi   \in
S({\Bbb R}^n)$,  в  частности,  любую  функцию  $\varphi  \in  D({\Bbb
R}^n)$, можно  приблизить  по  норме  $\Cal  H$  линейной  комбинацией
\r{3.33}. Поскольку   множество  $D({\Bbb  R}^n)$  плотно  в  $S({\Bbb
R}^n)$, а скалярное произведение \r{2.5} непрерывно согласно лемме  2.2
по своим   аргументам  в  топологии  $S({\Bbb  R}^n)$,  любой  элемент
$S({\Bbb R}^n,{\Cal L}_k)$ можно аппроксимировать функциями  \r{3.33}.
Отсюда получаем утверждение леммы.

Из лемм 3.13 и 3.14 вытекает первое утверждение теоремы 3.3.

{\bf Лемма 3.15.} {\it Любой элемент $\Cal H$ вида
$e^{i\Omega(Y)}f$, $Y\in {\Cal L}_k^{<\perp>}$,  может  быть  с  любой
точностью аппроксиморован элементами $\Cal H$ вида
\bez
[\Omega(Y_+)]^m f, \qquad Y_+\in \check{r}^*(A).
\eez
}

{\it Доказательство.} По первому утверждению теоремы 3.2 для любого $Y
\in {\Cal L}_k^{<\perp>}$ определены $Y_-\in \check{r}(A)$ и $Y_+  \in
\check{r}^*(A)$, такие, что $Y=Y_-+Y_+$, $Y_-=Y_+^*$. Имеем:
\bez
e^{i\Omega(Y)} f =
e^{i\Omega(Y_+)} e^{i\Omega(Y_-)} e^{-\frac{i}{2}<Y_+,Y_->} f
= c e^{i\Omega(Y_+)}f,
\eez
где $c = e^{-\frac{i}{2} <Y_+,Y_->}$. При этом ряд для экспоненты
\bez
e^{i\Omega(Y_+)}f = \sum_{n=0}^{\infty} \frac{i^n}{n!} (\Omega(Y_+))^n
f
\eez
сходится по норме, поскольку квадрат нормы $n$-го слагаемого равен
\bez
\frac{1}{n!} (\frac{1}{i} <Y_+^*,Y_+>)^n ||f||^2.
\eez
Отсюда получаем утверждение леммы и второе утверждение теоремы 3.3.

{\it Доказательство второго утверждения теоремы 3.2.}
Пусть две  гауссовские  функции $f_1$ и $f_2$ эквивалентны:  $f_1 \sim
f_2$. Поскольку оператор $\Omega(Y)$ при $f\in  {\Cal  L}_k^{<\perp>}$
переводит эквивалентные функции в эквивалентные, свойства
$\Omega(Y) f_1  \sim  0$  и  $\Omega(Y)  f_2  \sim   0$   равносильны.
Следовательно, $\check{r}(A_1) = \check{r}(A_2)$.

Пусть, обратно,  $\check{r}(A_1)  =  \check{r}(A_2) = \check{r}$.  Это
означает, что свойства
$\Omega(Y) f_1 \sim 0$ и $\Omega(Y) f_2 \sim 0$ равносильны при $Y \in
{\Cal L}_k^{<\perp>}$. Рассмотрим функцию
\bez
\tilde{f}_1 = f_1 - f_2 \frac{((f_2,f_1))}{((f_2,f_2))}.
\eez
Очевидно, $((\tilde{f}_1,f_2))=0$; кроме того,
\bey
((\tilde{f}_1, \Omega(Y_1^*) ... \Omega(Y_k^*) f_2)) =
((\Omega(Y_1) \tilde{f}_1,
\Omega(Y_2^*) ... \Omega(Y_k^*) f_2)) = ... = 0, \\
Y_1,...,Y_k \in \check{r}.
\eey
Следовательно, функция $\tilde{f}_1$ ортогональна всем функциям
$\Omega(Y_1^*) ... \Omega(Y_k^*) f_2$ и равна нулю в силу полноты этой
системы. Второе утверждение теоремы 3.2 полностью доказано.

{\it Доказательство леммы 3.7.}
Не ограничивая  общности,  можно  считать,  что  $Z  \in r_{\perp}(A)$,
поскольку $\omega(Z) g = 0$ при $Z \in {\Cal L}_k^{\Bbb C}$. Введем на
$r_{\perp}(A)$ ортонормированный  относительно скалярного произведения
$\frac{1}{i}<Z,Z^*>$ базис  ${\Cal   Y}^{(1)},...,{\Cal   Y}^{(n-k)}$,
одним из   элементов   которого   является  вектор  ${\Cal  Y}^{(1)}$,
пропорциональный $Z$. Введем ростковые операторы рождения и уничтожения
\bez
A_{\alpha}^+ = \Omega({\Cal Y}^{(\alpha)*},
\qquad
A_{\alpha}^- = \Omega({\Cal Y}^{(\alpha)},
\eez
удовлетворяющие каноническим коммутационным соотношениям
\beq
[A^{\pm}_{\alpha}, A^{\pm}_{\beta}]=0,
\qquad
[A^-_{\alpha},A^+_{\beta}] = \delta_{\alpha\beta}
\l{3.34}
\eeq
ввиду ортонормированности базиса. Утверждение леммы означает, что
\beq
||A_1^-g|| \le \sqrt{s} ||g||, \qquad
||A_1^+g|| \le \sqrt{s+1} ||g||
\l{3.35}
\eeq
при
\bez
g =               \sum_{p=0}^s              \sum_{\alpha_1...\alpha_p}
g^{(p)}_{\alpha_1...\alpha_p} A^+_{\alpha_1} ... A^+_{\alpha_p} [f],
\quad A_{\beta}^- [f] = 0.
\eez
Поскольку
\bey
||A_1^-g||^2 = ((g,A_1^+A_1^-g)) \le s((g,g)) = s||g||^2;\\
||A_1^+g||^2 = ||g||^2 + ||A_1^-g||^2 \le
(s+1)||g||^2.
\eey
Свойство \r{3.35} проверено, лемма 3.7 доказана.

\section{Квадратичные гамильтонианы}

\subsection{Формулировка результатов}

Эволюция квантовой   системы   задается   однопараметрической  группой
$e^{-iHt}$ унитарных   операторов,    действующих    в    гильбертовом
пространстве. Генератор  группы  $H$  называется оператором Гамильтона
(гамильтонианом).

В приложениях часто встречается важный частный случай  оператора  $H$,
квадратичного по   операторам   $\xi_i$   и  $\frac{\partial}{\partial
\xi_i}$. Выбирая  в  $\Cal  M$  базис  $Z^{(1)},...,Z^{(2n)}$,   можно
записать квадратичный оператор $H$ как
\beq
H =    \frac{1}{2}    \sum_{ij=1}^{2n}   \Gamma_{ij}
\Omega(Z^{(i)})
\Omega(Z^{(j)}) + {\varepsilon},
\l{4.1}
\eeq
где $\Gamma_{ij}$   ---   симметричная  матрица,  ${\varepsilon}$  ---
оператор умножения на число.

Выражение \r{4.1} можно записать и в более абстрактном виде. Обозначим
через $\Omega_2$ отображение, сопоставляющее элементу
\bez
\Gamma =  \frac{1}{2}  \sum_{ij=1}^{2n}  \Gamma_{ij}  Z^{(i)}  \otimes
Z^{(j)} \in Sym {\Cal M} \otimes {\Cal M}
\eez
оператор
\bez
\Omega_2(\Gamma) = \frac{1}{2} \sum_{ij=1}^{2n} \Gamma_{ij}
\Omega(Z^{(i)}) \Omega(Z^{(j)}).
\eez
Тем самым
\beq
H = \Omega_2(\Gamma) + {\varepsilon},  \qquad \Gamma \in Sym {\Cal  M}
\otimes {\Cal M}.
\l{4.2}
\eeq
Выражение \r{4.2} определяет оператор,  действующий в $S({\Bbb R}^n)$.
Чтобы оно определяло оператор в $\Cal H$, нужно потребовать, чтобы оно
сохраняло отношение эквивалентности.

{\bf Лемма 4.1.} {\it
1. Свойство
\bez
f_1 \sim f_2 \quad \Rightarrow \quad Hf_1 \sim Hf_2
\eez
справедливо для оператора \r{4.2} тогда и только тогда, когда
\bez
\Gamma \in Sym {\Cal L}_k^{<\perp>} \otimes {\Cal L}_k^{<\perp>} +
Sym {\Cal L}_k \otimes {\Cal M}.
\eez
2. Для любого оператора вида \r{4.2} можно  подобрать  такие
$\Gamma'
\in Sym   {\Cal   L}_k^{<\perp>}   \otimes   {\Cal  L}_k^{<\perp>}$
и ${\varepsilon}'$, что при всех $f\in S({\Bbb R}^n,{\Cal L}_k)$
\bez
Hf \sim [\Omega_2(\Gamma') + {\varepsilon}']f.
\eez
}

Доказательство приведено в пункте 4.2.

Как вытекает из леммы 4.1,  можно рассматривать только такие операторы
\r{4.2}, для которых
$\Gamma
\in Sym {\Cal  L}_k^{<\perp>}  \otimes  {\Cal  L}_k^{<\perp>}$.  Будем
рассматривать оператор \r{4.2} на множестве квазигауссовских векторов,
являющихся линейными комбинациями векторов вида
\bez
\Omega(Y_1^*)...\Omega(Y_s^*) [f]
\eez
для некоторой гауссовской функции $f$
\beq
f(\xi) = c \exp[\frac{i}{2} \sum_{ij} \xi_i A_{ij} \xi_j],  \quad Im A
> 0.
\l{4.3}
\eeq

{\bf Лемма   4.2.}   {\it  1.  Оператор  \r{4.2},  рассматриваемый  на
множестве квазигауссовских векторов, симметричен.

2. Квазигауссовские  векторы  являются  аналитическими  для  оператора
\r{4.2}.
}

Как вытекает  из   леммы   4.2,   замыкание   оператора   $H$   задает
самосопряженный оператор в $\Cal H$.

Для дальнейшего    введем    на   ${\Cal   L}_k^{<\perp>}$   отношение
эквивалентности. Именно,  будем говорить,  что два вектора $Y',Y'' \in
{\Cal L}_k^{<\perp>}$  эквивалентны  друг другу,  $Y' \sim Y''$,  если
$Y'-Y'' \in {\Cal L}_k$.  Обозначим соответствующее факторпространство
как
\bez
{\Cal R} = {\Cal L}_k^{<\perp>}/{\Cal L}_k.
\eez
Очевидно, что  кососкалярное  произведение  $<\cdot,\cdot>$  допускает
сужение на факторпространство.

Поскольку при $Y\in {\Cal L}_k$ справедливо свойство  $[\Omega(Y)f]  =
0$ при  всех  $f$,  каждому  $\overline{Y}  \in  {\Cal  R}$  однозначно
сопоставляется оператор    $\overline{\Omega}(\overline{Y})     \equiv
\Omega(Y)$, где $Y$ --- любой из представителей класса $\overline{Y}$.
Аналогично, элементу $\overline{\Gamma} \in Sym {\Cal R} \otimes {\Cal
R}$ однозначно     сопоставляется     оператор    $\overline{\Omega}_2
(\overline{\Gamma}) \equiv \Omega_2(\Gamma)$, где
$\Gamma \in Sym {\Cal L}_k^{<\perp>} \otimes {\Cal L}_k^{<\perp>}$ ---
любой из представителей класса эквивалентности $\overline{\Gamma}$.

При $\overline{\Gamma}     =
\frac{1}{2}\sum_{ij}      \Gamma_{ij}
\overline{Z}^{(i)} \otimes \overline{Z}^{(j)} \in Sym {\Cal R} \otimes
{\Cal R}$,  $\Gamma_{ij} = \Gamma_{ji}$,  $\overline{Y} \in {\Cal R}$
введем кососкалярное   произведение  $<\overline{Y},\overline{\Gamma}>
\in {\Cal R}$ следующим образом:
\bez
<\overline{Y}, \frac{1}{2} \sum_{ij}      \Gamma_{ij}
\overline{Z}^{(i)} \otimes \overline{Z}^{(j)}> \equiv
\sum_{ij} \Gamma_{ij}
<\overline{Y}, \overline{Z}^{(i)}> \overline{Z}^{(j)}.
\eez

Исследуем теперь свойства уравнения
\beq
i \frac{df(t)}{dt} \sim
[\overline{\Omega}_2(\overline{\Gamma}) + {\varepsilon}] f(t),
\l{4.4}
\eeq
где $f(t)  \in {\Cal H}$,  $\overline{\Gamma} \in Sym {\Cal R} \otimes
{\Cal R}$.

Оказывается, и для систем  со  связями  справедливо  свойство,  аналог
которого использовался при развитии теории комплексного ростка Маслова
для систем без связей.

{\bf Лемма 4.3.} {\it
Оператор $\overline{\Omega}  (\overline{Y}(t))$,  $\overline{Y}(t) \in
{\Cal R}$ коммутирует с оператором
$i\frac{d}{dt} -     \overline{\Omega}_2     (\overline{\Gamma})     -
{\varepsilon}$ на $S({\Bbb R}^n)$ тогда и только тогда, когда
\beq
\frac{d\overline{Y}(t)}{dt} =
<\overline{Y}(t), \overline{\Gamma}>.
\l{4.5}
\eeq
}

Уравнение \r{4.5}  на  $2n-2k$-мерный  вектор  $\overline{Y}(t)$ можно
также представить  как  систему   $2n-2k$   уравнений.   Эта   система
называется системой классических уравнений.

Лемма 4.3    показывает,    что    при    условии   \r{4.5}   оператор
$\overline{\Omega}(\overline{Y}(t))$ переводит    решение    уравнения
\r{4.4} в  решение.  Следовательно,  лемма 4.3 позволяет строить новые
решения уравнения \r{4.4} из известных решений.

Обозначим через $u_t$  разрешающий  оператор  для  уравнения  \r{4.5},
переводящий начальное  условие  для уравнения \r{4.5} в решение задачи
Коши для него в момент времени $t$:
\bez
u_t: \overline{Y}(0) \mapsto \overline{Y}(t).
\eez
Важным примером решения уравнения \r{4.4} является гауссовская функция
\beq
f(t,\xi) = c(t) e^{\frac{i}{2} \sum_{ij} \xi_i A_{ij}(t) \xi_j}.
\l{4.6}
\eeq
Ей соответствует функция $F=\eta f$ вида
\beq
F(t,\xi) = \check{c}(t)
e^{\frac{i}{2} \sum_{ij} \xi_i \check{A}_{ij}(t) \xi_j}.
\l{4.7}
\eeq

Обозначим через    $r(A(t))$    и   $\check{r}(A(t))$   соответственно
комплексный  $S$-росток  Маслова  и  комплексный  $H$-росток  Маслова,
отвечающие функции  \r{4.6}  в  момент $t$.  Через $C(t):  r(A(t)) \to
{\Bbb C}^n$ и $\check{C}(t): \check{r}(A(t)) \to {\Bbb C}^n$ обозначим
операторы проектирования  с  ростков на координатную плоскость,  через
${\Cal P}_-(t):  \check{r}(A(t))  \to  r(A(t))$  ---   оператор   вида
\r{3.10}.

{\bf Лемма 4.4.} {\it
Гауссовская функция \r{4.6} является решением уравнения \r{4.4}  тогда
и только тогда, когда выполнены два условия:

(а) $\check{r}(A(t)) = u_t \check{r}(A(0))$;

(б)
\beq
c(t) = c(0)
\frac{\Delta ({\Cal P}_-(t))}{\Delta ({\Cal P}_-(0))}
\frac{1}{\sqrt{det[C(t) {\Cal P}_-(t) u_t {\Cal P}^{-1}_-(0) C(0)]}}.
\l{4.8}
\eeq
При этом
\beq
\check{c}(t) =      \frac{\check{c}(0)}{\sqrt{det(\check{C}(t)     u_t
\check{C}^{-1}(0))}}.
\l{4.9}
\eeq
}

Из лемм 4.3 и 4.4 вытекает следствие.

{\bf Следствие.} {\it
Пусть $f(t,\xi)$ --- гауссовское решение уравнения \r{4.4}.
Тогда функция вида
\beq
\overline{\Omega}(u_t \overline{Y}_1) ...
\overline{\Omega}(u_t \overline{Y}_s) f(t,\xi)
\l{4.10}
\eeq
также является решением уравнения \r{4.4}.
}

Обобщим теперь   приведенные   результаты   на   случай,   когда   как
пространство ${\Cal L}_k$ (а значит, и гильбертово пространство ${\Cal
H} = {\Cal H}_t$),  так и коэффициенты $\Gamma_{ij}$ в  операторе  $H$
\r{4.1} зависят от времени $t$. Рассмотрим уравнение вида
\beq
i \frac{\partial f(t,\xi)}{\partial t} =
[\Omega_2(\Gamma(t)) + {\varepsilon}(t)] f(t,\xi);
\l{4.11}
\eeq
здесь $f(t,\cdot)   \in   S({\Bbb   R}^n)$,   производная  по  времени
и уравнение \r{4.11} понимаются в пространстве $S({\Bbb   R}^n)$.

Обозначим через $u_t$ отображение,  сопоставляющее начальному  условию
для уравнения
\beq
\frac{dY(t)}{dt} = <Y(t),\Gamma(t)>, \qquad Y(t) \in {\Cal M}
\l{4.12}
\eeq
в решение   задачи   Коши:   $u_t:Y(0)  \mapsto  Y(t)$.  Кососкалярное
произведение понимается в смысле
\bez
<Y, \sum_{ij} \frac{1}{2} \Gamma_{ij} Z^{(i)} \otimes Z^{(j)}>  \equiv
\sum_{ij} \Gamma_{ij} <Y,Z^{(i)}> Z^{(j)}.
\eez

{\bf Теорема  4.1.}  {\it  1.  Гауссовская  функция  \r{4.6}  является
решением уравнения \r{4.11} тогда и только тогда,  когда выполнены два
условия:

(а) $r(A(t)) = u_t r(A(0))$;

(б) $c(t) = c(0) \frac{
e^{-i\int_0^t dt' {\varepsilon}(t')}
}{\sqrt{det (C(t)u_t C^{-1}(0))}}$.

2. Пусть   начальное   условие   для   уравнения   \r{4.11}   является
квазигауссовской функцией.  Тогда  уравнение  \r{4.11} имеет решение в
классе квазигауссовских функций.

3. Преобразование  эволюции,   переводящее   начальное   условие   для
уравнения \r{4.11}  в  решение  задачи  Коши  в момент $t$,  сохраняет
отношение эквивалентности и  скалярное  произведение  тогда  и  только
тогда, когда
\beq
{\Cal L}_k(t) = u_t {\Cal L}_k(0).
\l{4.13}
\eeq
\beq
Im {\varepsilon} = \frac{1}{2} \frac{d}{dt} \ln \Delta (u_t^0),
\l{4.14}
\eeq
где $u_t^0:  {\Cal  L}_k(0)  \to  {\Cal  L}_k(t)$ --- сужение $u_t$ на
${\Cal L}_k(0)$.

4. При выполнении условия \r{4.13} справедливы свойства
\bez
\check{r}(A(t)) = u_t \check{r}(A(0)), \qquad
\check{c}(t) = \frac{\check{c}(0)
e^{-i\int_0^t Re {\varepsilon}(t') dt'} \sqrt{\Delta (u_t^0)}
}
{\sqrt{det[ \check{C}(t) u_t \check{C}^{-1}(0)]}}.
\eez
}

Как вытекает   из   пункта   2,  при  условии  \r{4.13}  на  множестве
квазигауссовских функций    определен     изометрический     оператор,
переводящий начальное  условие  для уравнения \r{4.11} в решение этого
уравнения. Ввиду   полноты   системы   квазигауссовских   функций   он
однозначно продолжается  до  изометрического оператора $U_{t0}:  {\Cal
H}_0 \to  {\Cal  H}_t$.  Поскольку  начальный  момент  можно   выбрать
произвольным образом,  определены  изометрические  операторы  эволюции
$U_{t_2t_1}: {\Cal  H}_{t_1}  \to  {\Cal  H}_{t_2}$,   удовлетворяющие
свойству $U_{t_3t_2}     U_{t_2t_1}     =    U_{t_3t_1}$.    Поскольку
$U_{t_0t_0}=1$, операторы $U_{t_2t_1}$ обратимы и унитарны.

\subsection{Доказательства утверждений}

{\it Доказательство леммы 4.1.}

Докажем достаточность в первом утверждении.  Пусть $f\sim 0$ и $\Gamma
\in Sym  {\Cal L}_k^{<\perp>} \otimes {\Cal L}_k^{<\perp>} + Sym {\Cal
L}_k \otimes {\Cal M}$. Покажем, что $Hf \sim 0$.
Заметим, что  операторы  $\Omega(Z)$,  $Z  \in  {\Cal  L}_k^{<\perp>}$
сохраняют отношение эквивалентности (лемма 2.5):  $\Omega(X) \Omega(Y)
f \sim 0$ при $Y \in  {\Cal  L}_k$  и  любом  $f$  (формула  \r{2.11}),
$\Omega(Y) \Omega(X)  f  =  \Omega(X) \Omega(Y) f - i <Y,X> f \sim 0$.
Отсюда $Hf \sim 0$.

Докажем необходимость. Воспользуемся результатом леммы 2.8:
\bez
{\Cal M} = {\Cal L}_k + {\Cal G}_k +
({\Cal L}_k + {\Cal G}_k)^{<\perp>}
\eez
и введем базисы
$({\Cal X}^{(1)},...,{\Cal X}^{(k)})$
на ${\Cal L}_k$ и $({\Cal Y}^{(1)},...,{\Cal Y}^{(k)})$
на ${\Cal G}_k$, удовлетворяющие свойству
\beq
<{\Cal X}^{(a)},{\Cal Y}^{(b)}> = \delta^{ab},
\l{4.15}
\eeq
и базис
$({\Cal Z}^{(1)},...,{\Cal Z}^{(2n-2k)})$
на $({\Cal L}_k + {\Cal G}_k)^{<\perp>}$. Тогда можно записать:
\beb
H = \frac{1}{2}
\sum_{ab} \Gamma^{(1)}_{ab} \Omega({\Cal X}^{(a)}) \Omega({\Cal X}^{(b)})
+ \frac{1}{2}
\sum_{\alpha\beta}
\Gamma^{(2)}_{\alpha\beta} \Omega({\Cal Y}^{(\alpha)})
\Omega({\Cal Y}^{(\beta)}) \\
+
\frac{1}{2}
\sum_{ij}
\Gamma^{(3)}_{ij} \Omega({\Cal Z}^{(i)})
\Omega({\Cal Z}^{(j)}) +
\frac{1}{2}
\sum_{a\alpha}
\Gamma^{(4)}_{a\alpha}
(\Omega({\Cal X}^{(a)})
\Omega({\Cal Y}^{(\alpha)}) +
\Omega({\Cal Y}^{(\alpha)})
\Omega({\Cal X}^{(a)}))
\\
+
\sum_{ia}
\Gamma^{(5)}_{ia} \Omega({\Cal X}^{(a)})
\Omega({\Cal Z}^{(i)}) +
\sum_{i\alpha}
\Gamma^{(6)}_{i\alpha} \Omega({\Cal Y}^{(\alpha)})
\Omega({\Cal Z}^{(i)}) +
{\varepsilon}.
\l{4.16}
\eeb
Здесь использованы свойства
\bez
[\Omega({\Cal X}^{(a)});\Omega({\Cal Z}^{(i)})] = 0,
\qquad
[\Omega({\Cal Y}^{(a)});\Omega({\Cal Z}^{(i)})] = 0.
\eez
Рассмотрим функцию $f\sim 0$ вида
\bez
f = \Omega({\Cal X}^{(c)} g \sim 0.
\eez
Тогда ввиду \r{4.15}
\beq
[\Omega({\Cal X}^{(a)});\Omega({\Cal Y}^{(b })] = -i \delta_{ab}
\l{4.17}
\eeq
и
\bez
Hf \sim
\sum_{\beta} \Gamma^{(2)}_{c\beta} (-i) \Omega({\Cal Y}^{(\beta)}) g
+ \sum_i \Gamma^{(6)}_{ic} \Omega({\Cal Z}^{(i)}) (-i) g.
\eez
Ввиду следствия из теоремы 3.1 отсюда вытекает, что
\bez
\sum_{\beta} \Gamma^{(2)}_{c\beta} {\Cal Y}^{(\beta)} +
\sum_c \Gamma^{(6)}_{ic} {\Cal Z}^{(i)} \in {\Cal L}_k,
\eez
или
\beq
\Gamma^{(2)}_{c\beta} = 0, \qquad
\Gamma^{(6)}_{ic} = 0.
\l{4.18}
\eeq
Свойство \r{4.18} как раз и означает, что
\bez
\Gamma \in Sym  {\Cal L}_k^{<\perp>} \otimes {\Cal L}_k^{<\perp>}
+ Sym {\Cal L}_k \otimes {\Cal M}.
\eez

Докажем теперь  второе  утверждение   леммы.   Воспользуемся   записью
\r{4.16} и формулой \r{2.11}. Тогда
\bez
Hf \sim
\left(
\frac{1}{2}
\sum_{ij}
\Gamma^{(3)}_{ij} \Omega({\Cal Z}^{(i)})
\Omega({\Cal Z}^{(j)}) +
\frac{1}{2}
\sum_{a\alpha}
\Gamma^{(4)}_{a\alpha}
[\Omega({\Cal Y}^{(\alpha)});
\Omega({\Cal X}^{(a)})]
+ {\varepsilon}
\right) f,
\eez
что доказывает утверждение ввиду \r{4.17}.

{\it Доказательство леммы 4.2.}

Симметричность оператора \r{4.2} вытекает из свойства \r{2.12}. Докажем
аналитичность квазигауссовских векторов. Требуется проверить, что ряд
\beq
\sum_{n=0}^{\infty} \frac{||H^ng||}{n!} t^n
\l{4.19}
\eeq
сходится при  достаточно  малых  $t$.  Пусть  $g$ --- квазигауссовский
вектор ранга $s$. Тогда из леммы 3.7 вытекает, что
\bez
||Hg|| \le \sqrt{(s+1)(s+2)} a||g||, \qquad a = const.
\eez
При этом $Hg$ --- квазигауссовский вектор ранга $s+2$. Следовательно,
\bez
||H^ng|| \le a^n \sqrt{(s+1)...(s+2n)} ||g|| = a^n
\frac{\sqrt{(s+2n)!}}{\sqrt{s!}} ||g||,
\eez
и ряд  \r{4.19},  как  вытекает  из  формулы  Стирлинга для $n!$,  при
достаточно малых $t$ сходится.
Утверждение доказано.

{\it Доказательство леммы 4.3.}
Непосредственным вычислением устанавливаем, что
\bez
\left[
i \frac{d}{dt}; \overline{\Omega}(\overline{Y}(t))
\right]
= i \overline{\Omega} (\dot{\overline{Y}}(t));
\quad
[ H; \overline{\Omega}(\overline{Y}) ] =
i \overline{\Omega}(<\overline{Y},\Gamma>).
\eez
Отсюда получаем, что условие
$\left[
i \frac{d}{dt}; \overline{\Omega}(\overline{Y}(t))
\right] = 0$ равносильно
$\overline{\Omega}(\dot{\overline{Y}} - <\overline{Y},\Gamma>)  =  0$,
что по следствию из теоремы 3.1 равносильно
$\dot{\overline{Y}} - <\overline{Y},\Gamma>  =  0$
в смысле факторпространства. Лемма 4.3 доказана.

{\it Доказательство леммы 4.4.}
Сначала докажем более простое утверждение.

{\bf Лемма 4.5}.  {\it 1.
Пусть  гауссовская  функция  \r{4.6}  является
решением уравнения   \r{4.4}.    Тогда    $\check{r}(A(t))    =    u_t
\check{r}(A(0))$. \newline
2. Пусть $\check{r}(A(t)) = u_t\check{r}(A(0))$.  Тогда для некоторого
числового множителя  $c(t)$ гауссовская функция \r{4.6} будет являться
решением уравнения \r{4.4}.
}

{\it Доказательство.}  Для  доказательства  первого  утверждения леммы
заметим, что функция  $\overline{\Omega}(u_t\overline{Y})[f(t,\cdot)]$
удовлетворяет уравнению  \r{4.4}  для  любого  $\overline{Y} \in {\Cal
R}$. Следовательно,  если она равна нулю в один момент времени $t$, то
она равна нулю и в остальные моменты. Таким образом, свойства
\bez
\Omega(u_tY) f(t,\cdot) \sim 0, \qquad
\Omega(Y) f(0,\cdot) \sim 0
\eez
равносильны, что и доказывает первое утверждение.

Докажем второе утверждение.  Пусть $f_t = e^{-iHt}  f(0,\cdot)$
--- решение  уравнения  \r{4.4}  с  заданными  гауссовскими начальными
условиями. Проверим,  что оно совпадает с $f(t,\cdot)$ с точностью  до
числового множителя.  Ввиду  полноты  системы квазигауссовских функций
достаточно проверить, что
\bez
((\Omega(Y_1^*) ... \Omega(Y_p^*) f(t,\cdot), f_t)) = 0, \qquad
Y_i \in \check{r}(A(t)),
\eez
или
\beq
\Omega(Y_t) f_t \sim 0, \qquad Y_t \in \check{r}(A(t)).
\l{4.20}
\eeq
Но свойство \r{4.20} равносильно соотношению
\bez
\Omega(Y_0) f_0 \sim 0, \qquad Y_0 \in \check{r}(A_0),
\eez
вытекающему непосредственно из определения ростка. Лемма 4.5 доказана.

Докажем теперь формулы \r{4.8} и \r{4.9}.  Прежде всего,  заметим, что
ввиду леммы 3.6 они равносильны. Поэтому ограничимся проверкой формулы
\r{4.9}. Запишем оператор $H$ как
\bey
H =
\frac{1}{2} \sum_{ij=1}^n
(-i \frac{\partial}{\partial \xi_i})
H_{P_iP_j}
(-i \frac{\partial}{\partial \xi_j})
+ \\
\frac{1}{2} \sum_{ij=1}^n
H_{P_iQ_j} [(-i \frac{\partial}{\partial \xi_i}) \xi_j
+ \xi_j (-i \frac{\partial}{\partial \xi_i})]
+
\frac{1}{2} \sum_{ij=1}^n
H_{Q_iQ_j} \xi_i\xi_j + {\varepsilon}.
\eey
Если $f$  удовлетворяет  уравнению \r{4.4},  то обобщенная функция $F$
\r{4.7} будет удовлетворять уравнению
\bez
i \frac{\partial F(t,\xi)}{\partial t} = HF(t,\xi).
\eez
Отсюда получаем уравнения на $\check{c}(t)$ и $\check{A}(t)$:
\beb
i\frac{d}{dt} \check{c}(t) =
\left[
\sum_{ij=1}^n H_{P_iP_j} \check{A}_{ji} +  \sum_{i=1}^n  H_{P_iQ_i}  +
{\varepsilon}
\right] \check{c}(t);\\
i\frac{d}{dt} \check{A} = H_{PP} + H_{PQ}\check{A} + \check{A}H_{QP} +
\check{A}H_{PP}\check{A}.
\l{4.21}
\eeb
При этом классическая система \r{4.12} для
$Y(t) = \left(\matrix P(t) \\ Q(t) \endmatrix\right)$ запишется как
\bey
\frac{d}{dt}P = - H_{QP} P - H_{QQ} Q; \\
\frac{d}{dt}Q = - H_{PP} P - H_{PQ} Q.
\eey
Такой же по форму системе удовлетворяют и матрицы
\bez
\tilde{B}(t) = \check{B}(t) u_t \check{C}^{-1}(0),
\quad
\tilde{C}(t) = \check{C}(t) u_t \check{C}^{-1}(0).
\eez
Именно,
\bey
\frac{d}{dt}\tilde{B} = - H_{QP} \tilde{B} - H_{QQ} \tilde{C}; \\
\frac{d}{dt}\tilde{C} = - H_{PP} \tilde{B} - H_{PQ} \tilde{C}.
\eey
При этом  $\check{A}  =   \tilde{B}   \tilde{C}^{-1}$,   а   $ln   det
\tilde{C}(t)$ удовлетворяет уравнению
\bez
\frac{d}{dt}[ln det    \tilde{C}(t)]    =   Tr   \frac{d\tilde{C}}{dt}
\tilde{C}^{-1} = Tr [H_{PP} \check{A} + H_{PQ}].
\eez
Отсюда получаем утверждение леммы 4.4.

{\it Доказательство теоремы 4.1.}

Первое утверждение доказывается по аналогии с леммой  4.4.  Решение  в
классе квазигауссовских  функций имеет вид линейной комбинации функций
\r{4.10}.

Проверим третье утверждение.  Пусть преобразование эволюции  сохраняет
отношение эквивалентности.
Пусть $f(t,\xi) = g_t(\xi)$  ---  квазигауссовское  решение  уравнения
\r{4.11}. Тогда начальное состояние вида
\beq
\Omega(X_0)g_0 \sim 0, \qquad X_0 \in {\Cal L}_k(0)
\l{4.22}
\eeq
переходит в момент $t$ в состояние вида
\beq
\Omega(u_tX_0) g_t \sim 0,
\l{4.23}
\eeq
которое также  должно  быть  эквивалентно  нулю  для любой $g_t$.  Это
означает, что  $u_tX_0  \in  {\Cal  L}_k(t)$.  Обратно,  из   свойства
\r{4.23} вытекает  \r{4.22},  что  означает ${\Cal L}_k(t) = u_t {\Cal
L}_k(0)$.

Пусть условие \r{4.13} выполнено. Покажем, что норма сохраняется тогда
и только  тогда,  когда  выполнено  соотношение  \r{4.14}.  Для этого
выберем на ${\Cal L}_k(t)$ базис ${\Cal  X}^{(a)}(t)$  таким  образом,
чтобы
\beq
{\Cal X}^{(a)}(t) = u_t {\Cal X}^{(a)}(0);
\l{4.24}
\eeq
тогда
\bez
\eta(t) = J(t) \int d\alpha_1...d\alpha_k
e^{i\sum_a \alpha_a
\Omega({\Cal X}^{(a)}(t))},
\eez
преобразованию $u_t$ соответствует единичная матрица и
\bez
\Delta(u_t^0) = J(t)/J(0).
\eez
Из свойства \r{4.24} вытекает, что операторы
$\Omega({\Cal X}^{(a)}(t))$
и
$e^{i\sum_a \alpha_a \Omega({\Cal X}^{(a)}(t))}$
коммутируют с
$i\frac{d}{dt} - \Omega_2(\Gamma) - {\varepsilon}$;
следовательно,
\bey
i \frac{d}{dt} ((f(t,\cdot),f(t,\cdot)) =
i \frac{d}{dt} (f,\eta f) =
\\
([-\Omega_2(\Gamma) - {\varepsilon}]f,\eta f) +
(f, i\frac{d\eta}{dt} f)
+ (f,\eta [\Omega_2(\Gamma) + {\varepsilon}]f) =
\\
(f,[i\frac{d}{dt} - \Omega_2(\Gamma),\eta]f)
+ ({\varepsilon} - {\varepsilon}^*) (f,\eta f) =
\\
(f,i\frac{dln J}{dt} \eta f) + 2i Im{\varepsilon} (f,\eta f).
\eey
Таким образом, квадрат нормы сохраняется при условии
\bez
Im {\varepsilon} = - \frac{1}{2} \frac{dln J}{dt} =
- \frac{1}{2} \frac{d ln \Delta(u_t^0)}{dt},
\eez
что соответствует \r{4.14}.

Для проверки  четвертого  утверждения  получим уравнение на обобщенную
функцию $F$. Имеем:
\bey
[ i \frac{d}{dt} - \Omega_2(\Gamma) - Re {\varepsilon} ] F =
[ i \frac{d}{dt} - \Omega_2(\Gamma); \eta ] F +
\eta
[ i \frac{d}{dt} - \Omega_2(\Gamma) - Re {\varepsilon} ] f =
\\
i \frac{d}{dt} ln J(t) F -
\frac{i}{2} \frac{d}{dt} ln J(t) F =
\frac{i}{2} \frac{dln \Delta (u_t^0)}{dt} F.
\eey
Отсюда по аналогии с леммой 4.4 получаем утверждение леммы.

\section{Гауссовские собственные функции и устойчивость классической системы
уравнений}

Оказывается, что  вопрос  о  существовании   гауссовской   собственной
функции квадратичного гамильтониана
\beq
H = \overline{\Omega}_2(\overline{\Gamma}),
\qquad
\overline{\Gamma} \in Sym {\Cal R} \otimes {\Cal R},
\l{5.1}
\eeq
связан с вопросом об устойчивости классической системы \r{4.5}:
\beq
\frac{d\overline{Y}(t)}{dt} = <\overline{Y}(t),\overline{\Gamma}>,
\qquad \overline{Y}(t) \in {\Cal R}.
\l{5.2}
\eeq
Напомним, что система  \r{5.2}  называется  устойчивой,  если  все  ее
решения ограничены равномерно по $t$.

Сформулируем утверждение,  обобщающее теорему Маслова [11]
на случай систем
со связями.

{\bf Теорема 5.1}. {\it
1. Пусть  оператор  \r{5.1}  имеет гауссовский собственный вектор $[f]
\in {\Cal H}$, $f$ имеет вид \r{3.1}. Тогда система \r{5.2} устойчива.
\newline
2. Пусть система \r{5.2} устойчива. Тогда:\newline
(а) существует  $2n-2k$ линейно-независимых решений уравнения \r{5.2},
которые разбиваются на пары сопряженных решений
\beq
\overline{Y}^{(I)}(t) = \overline{Y}^{(I)} e^{i\beta_It},
\quad
\overline{Y}^{(I)*}(t) = \overline{Y}^{(I)*} e^{-i\beta_It},
\quad
I = \overline{1,n-k}
\l{5.3}
\eeq
и удовлетворяет свойствам
\bez
\frac{1}{i} <\overline{Y}^{(I)},\overline{Y}^{(J)*}>                 =
\delta_{IJ},\quad
<\overline{Y}^{(I)},\overline{Y}^{(J)}> = 0;
\eez
(б) линейная оболочка $\check{r} \subset {\Cal M}^{\Bbb C}$, натянутая
на подпространство     ${\Cal     L}_k^{\Bbb     C}$     и     векторы
$Y^{(1)},...,Y^{(n-k)}$, являющиеся       представителями      классов
эквивалентности $\overline{Y}_1,...,\overline{Y}^{(n-k)}$,  не зависит
от выбора    представителей   классов   эквивалентности   и   является
комплексным $H$-ростком Маслова для некоторой гауссовской функции  $f$
\r{3.1}: $\check{r} = \check{r}(A)$;
\newline
(в) векторы вида
\beq
\overline{\Omega}(\overline{Y}^{(1)*})^{N_1}
...
\overline{\Omega}(\overline{Y}^{(n-k)*})^{N_{n-k}}
[f]
\l{5.4}
\eeq
образуют полную  ортогональную  систему собственных векторов оператора
\r{5.1} с собственными значениями
\beq
\beta_1 (N_1 + \frac{1}{2}) + ...
+ \beta_{n-k} (N_{n-k} + \frac{1}{2}).
\l{5.5}
\eeq
}

Подчеркнем, что устойчивость системы \r{5.2} подразумевается именно  в
смысле факторпространства $\Cal R$, а не исходного пространства ${\Cal
L}_k^{<\perp>}$. Проиллюстрируем этот факт на примере.

Пусть
\bez
H= \xi_1\xi_2,
\eez
а пространство ${\Cal L}_1$ одномерно и натянуто на вектор
$(P_1=1,P_2=0,Q_1=0,Q_2=0)$. Очевидно,   что   $H\sim   0$   и   любая
гауссовская функция является собственной для $H$.

Исследуем устойчивость классической системы.  В данном  случае  ${\Cal
L}_1^{<\perp>}$ состоит  из  всех  векторов,  для  которых $Q_1=0$,  а
система \r{4.12} на ${\Cal L}_1^{<\perp>}$ имеет вид
\bey
\dot{P}_1 = - Q_2, \quad \dot{P}_2 = - Q_1;\\
\dot{Q}_2 = 0, \quad \dot{Q}_1 = 0;
\eey
ее решение линейно растет со временем
\bey
Q_1(t) = 0, \quad P_1(t) = P_{10} - Q_{20} t,
\\
Q_2(t) = Q_{20}, \quad P_2(t) = P_{20}.
\eey
Однако на факторпространстве решение данной системы постоянно, так как
разность векторов \newline
$(P_1(t),P_2(t),Q_1(t),Q_2(t))$
в  моменты  времени
$t_1$ и $t_2$ принадлежит ${\Cal L}_1$. Следовательно, система \r{5.2}
для факторпространства ${\Cal L}_1^{<\perp>}/{\Cal L}_1$ устойчива,  в
отличие от системы для ${\Cal L}_1^{<\perp>}$.

{\it Доказательство теоремы 5.1.}

Докажем первое  утверждение.  Пусть  $[f]$ --- гауссовский собственный
вектор гамильтониана \r{5.1},  а $\overline{Y}(t)$ --- решение системы
\r{5.2}. Покажем, что оно ограничено.

Пусть $Y  \in  {\Cal L}_k^{<\perp>}$;  тогда $Y$ по теоремам 3.1 и 3.2
можно представить в виде
\beq
Y = Y_- + Y_-^*,
\l{5.6}
\eeq
где $Y_-\in \check{r}(A)$,  $Y_-^* \in \check{r}(A)$.  При этом вектор
$Y_-$ определен однозначно с точностью до добавления вектора из ${\Cal
L}_k^{\Bbb C}$. Запишем
\bez
Y_- = Y_-^{\parallel} + Y_-^{\perp},
\quad
Y_-^{\perp} \in r_{\perp}(A), \quad
Y_-^{\parallel} \in {\Cal L}_k^{\Bbb C}.
\eez
Следовательно, вектор $Y_-^{\perp}  \in  r_{\perp}(A)$  сопоставляется
вектору $Y$ однозначно.  Данное отображение не меняется при добавлении
к $Y$ элементов  из  ${\Cal  L}_k$  и  допускает  поэтому  сужение  на
факторпространство
\bez
R_{\perp}: \overline{Y} \to Y_-^{\perp}.
\eez
Отображение $R_-$    действует    из   $2n-2k$-мерного   вещественного
пространства на $n-k$-мерное комплексное пространство,  которое  можно
рассматривать как $2n-2k$-мерное вещественное.

Покажем, что  $R_{\perp}$  является  взаимно однозначным отображением.
Предположим противное.  Тогда $R_{\perp}\overline{Y}=0$ для некоторого
$\overline{Y}\ne 0$.  Это  означает,  что  некоторый представитель $Y$
класса $\overline{Y}$ раскладывается в сумму  \r{5.6}  при  $Y_-,Y_-^*
\in {\Cal   L}_k$.   Следовательно,   $Y\in   {\Cal  L}_k$.  Но  тогда
$\overline{Y}=0$. Поэтому $R_{\perp}$ взаимно однозначно.

Проверим ограниченность  решения   $\overline{Y}(t)$.   Заметим,   что
квадрат нормы вектора
\bez
\Omega(\overline{Y}(t)) [f] =
\Omega(R_{\perp} \overline{Y}(t)) [f] +
\Omega((R_{\perp} \overline{Y}(t))^*) [f] =
\Omega((R_{\perp} \overline{Y}(t))^*) [f],
\eez
равная
\bez
\frac{1}{i}
<R_{\perp}\overline{Y}, (R_{\perp}\overline{Y})^*> =
|R_{\perp}\overline{Y}|^2
\eez
сохраняется. Поскольку  $R_{\perp}^{-1}$  ---  ограниченный   оператор
ввиду конечномерности  и  взаимной  однозначности,  это означает,  что
$\overline{Y}(t)$ ограничено в любой из норм в ${\Bbb R}^{2n-2k}$.
Первое утверждение теоремы доказано.

Докажем утверждение  2а.  Из теоремы о приведении матрицы к жордановой
нормальной форме  и  устойчивости  вытекает,  что  существует  $2n-2k$
линейно независимых    решений   системы   \r{5.2},   проворциональных
$e^{i\beta_{\alpha}t}$, $\alpha= \overline{1,2n-2k}$,  $\beta_{\alpha}
\in {\Bbb R}$. Это значит, что
\beq
i\beta_{\alpha} \overline{Y}^{(\alpha)} =
<\overline{Y}^{(\alpha)},\overline{\Gamma}>.
\l{5.7}
\eeq

Введем на $\Cal R$ билинейную форму
\beq
\nu(Y,Y') = \frac{1}{i} <Y,Y'{}^*>.
\l{5.8}
\eeq
Покажем, что  собственные  векторы  задачи \r{5.7},  отвечающие разным
$\beta$, ортогональны друг другу относительно скалярного  произведения
\r{5.8}. Действительно, пусть
\bez
i\beta \overline{Y} = <\overline{Y},\overline{\Gamma}>,
\quad
-i\beta' \overline{Y}'{}^* = <\overline{Y}'{}^*,\overline{\Gamma}>,
\quad \beta \ne \beta'.
\eez
Воспользуемся свойством
\bez
<\overline{Y}, <\overline{Y}'{}^*,\overline{\Gamma}>> =
<\overline{Y}'{}^*, <\overline{Y},\overline{\Gamma}>>,
\eez
вытекающим из симметричности $\overline{\Gamma}$, и запишем его в виде
\bez
i\beta <\overline{Y}'{}^*,\overline{Y}> =
- i\beta' <\overline{Y},\overline{Y}'{}^*>.
\eez
Отсюда
\bez
\nu(Y,Y') = 0.
\eez
Обозначим через ${\Cal R}_{\beta}$ собственное подпространство  задачи
\r{5.7}, отвечающее    собственному    значению   $\beta$.   Скалярное
произведение \r{5.8} на ${\Cal R}_{\beta}$ невырождено. Действительно,
предположим, что имеется вектор $Y\in {\Cal R}_{\beta}$, ортогональный
${\Cal R}_{\beta}$.  Тогда $Y$  будет  ортогонален  и  всем  остальным
подпространствам ${\Cal  R}_{\beta'}$,  а  значит,  и ${\Cal R}$,  что
противоречит невырожденности формы \r{5.8} на ${\Cal R}$.

Приведем форму \r{5.8} к каноническому виду на  ${\Cal  R}_{\beta}$  и
разложим ${\Cal  R}_{\beta}$ в прямую сумму ортогональных относительно
индефинитного скалярного произведения \r{5.8} подпространств
\bez
{\Cal R}_{\beta} = {\Cal R}_{\beta}^+ + {\Cal R}_{\beta}^-;
\eez
при этом на ${\Cal R}_{\beta}^+$ индефинитное  скалярное  произведение
положительно определено,  на  ${\Cal  R}_{\beta}^-$  ---  отрицательно
определено. При   этом,   поскольку   ${\Cal   R}_{-\beta}   =   {\Cal
R}_{\beta}^*$, разложение   пространства   ${\Cal  R}_{-\beta}$  можно
выбрать в виде
\bez
{\Cal R}_{-\beta} =
({\Cal R}_{\beta}^+)^* + ({\Cal R}_{\beta}^-)^*
\eez
на $({\Cal   R}_{\beta}^+)^*$   скалярное   произведение  отрицательно
определено; на $({\Cal R}_{\beta}^-)^*$ ---  положительно  определено.
Разложение же пространства ${\Cal R}_0 = {\Cal R}_0^*$ выберем в виде
\bez
{\Cal R}_0 = {\Cal R}_0^+ + ({\Cal R}_0^+)^*.
\eez
Выберем в   пространстве   $\sum  {\Cal  R}_{\beta}^+  +  \sum  ({\Cal
R}_{\beta}^-)^* +    {\Cal     R}_0^+$     ортонормированный     базис
$\{\overline{Y}^{(I)}\}$. Он   будет   удовлетворять   всем  требуемым
свойствам, что доказывает утверждение 2а.

Утверждение 2б является следствием теоремы 3.2.

Свойство полноты системы векторов \r{5.4} вытекает из теоремы 3.3.
Ортогональность этой   системы  векторов  вытекает  из  коммутационных
соотношений
\beq
[\overline{\Omega}(\overline{Y}^{(I)})
\overline{\Omega}(\overline{Y}^{(J)*}) ] = \delta_{IJ},
\qquad
[\overline{\Omega}(\overline{Y}^{(I)})
\overline{\Omega}(\overline{Y}^{(J)}) ] = 0,
\l{5.9}
\eeq
являющихся следствиями \r{5.3}.

Покажем теперь,   что   векторы  \r{5.4}  являются  собственными  для
оператора \r{5.1}. Для этого установим сначала, что
\beq
\overline{\Gamma} = \frac{1}{2} \sum_I
\beta_I [
\overline{Y}^{(I)*} \otimes \overline{Y}^{(I)}
+ \overline{Y}^{(I)} \otimes \overline{Y}^{(I)*}]
\l{5.10}
\eeq
Рассмотрим разность
\bez
\tilde{\Gamma} = \overline{\Gamma} -
\frac{1}{2}
\sum_I \beta_I [
\overline{Y}^{(I)*} \otimes \overline{Y}^{(I)}
+ \overline{Y}^{(I)} \otimes \overline{Y}^{(I)*}]
\eez
Тогда ввиду \r{5.7}
\bez
<\overline{Y}^{(J)},\tilde{\Gamma}> = i\beta_J \overline{Y}^{(J)} -
\sum_I \beta_I                <\overline{Y}^{(J)},\overline{Y}^{(I)*}>
\overline{Y}^{(I)} = 0,
\eez
аналогично получим, что
\bez
<\overline{Y}^{(J)*},\tilde{\Gamma}> = 0.
\eez
Ввиду свойства                     полноты                     системы
$\{\overline{Y}^{(J)},\overline{Y}^{(J)*}\}$ отсюда получаем
\bez
<\overline{Y},\tilde{\Gamma}> = 0, \qquad \overline{Y} \in {\Cal R}
\eez
и $\tilde{\Gamma}    =   0$   ввиду   невырожденности   кососкалярного
произведения на  ${\Cal  R}$.  Свойство  \r{5.10}  проверено;  из  него
вытекает, что
\beb
H = \frac{1}{2} \sum_I \beta_I
[
\overline{\Omega}(\overline{Y}^{(I)*})
\overline{\Omega}(\overline{Y}^{(I)})
+
\overline{\Omega}(\overline{Y}^{(I)})
\overline{\Omega}(\overline{Y}^{(I)*})
]
\\ =
\sum_I \beta_I
[\overline{\Omega}(\overline{Y}^{(I)*})
\overline{\Omega}(\overline{Y}^{(I)})
+ \frac{1}{2}
]
\l{5.11}
\eeb
Из коммутационных   соотношений   \r{5.9}  непосредственным  расчетом
получаем, что векторы  \r{5.4}  является  собственным  для  оператора
\r{5.11} с собственными значениями \r{5.5}.

Работа выполнена    при   финансовой   поддержке   Российского   фонда
фундаментальных исследований, проекты 08-01-00601-а и 05-01-02807-НЦНИЛ.

\newpage

СПИСОК ЛИТЕРАТУРЫ

1. {\it Дирак П.А.М.} Лекции по квантовой механике. М.:Мир, 1968.

2. {\it Арнольд В.И.} Математические методы классической механики. М.:Наука,
1989.

3. {\it Славнов А.А., Фаддеев Л.Д.} Введение в квантовую теорию калибровочных
полей. М.:Наука, 1988.

4. {\it Рубаков В.А.} Классические калибровочные поля. М.:УРСС, 1999.

5. {\it Ландау Л.Д.,  Лифшиц Е.М. }  Квантовая   механика.   Нерелятивистская
теория. М.:Наука, 1989.

6. {\it  фон  Нейман  И.}  Математические  основы  квантовой механики.
М.:Наука, 1964.

7. {\it Ashtekar A., Lewandowski J.,Marolf D., Mourao J.  and
     Thiemann T.}// J. Math. Phys., 1995, vol. 36, p.6456.

8. {\it Giulini D., Marolf D. }// Class.Q.Grav. 1999. v.16. p.2489;

9. {\it Shvedov O.Yu.}//Ann.Phys. 2002. v.302. p.2.

10. {\it Маслов В.П.} Операторные методы. М.:Наука, 1973.

11. {\it Маслов В.П.} Комплексный метод ВКБ в неличнейных  уравнениях.
М.:Наука, 1977.

12. {\it  Шведов О.Ю.} Метод комплексного ростка Маслова для систем со
связями первого рода.//ТМФ, 2003, т.136. с.418.

13. {\it Маслов В.П.,  Шведов О.Ю.} Метод комплексного ростка в задаче
многих частиц и квантовой теории поля. М.:УРСС, 2000.

\end{document}